\documentclass[%
 reprint,
 superscriptaddress,
 amsmath,amssymb,
 aps,
 prl,
]{revtex4-2}

\usepackage{graphicx}
\usepackage{dcolumn}
\usepackage{bm}
\usepackage{enumitem}   
\usepackage{hyperref}
\usepackage{url}
\usepackage{siunitx}
\usepackage{color}
\usepackage[normalem]{ulem}
\usepackage{makecell}
\usepackage{xcolor}

\begin{document}

\title{{Non-affine mechanics of entangled networks inspired by intermediate filaments}}

\author{Marco Pensalfini}
\email[Email: ]{marco.pensalfini@upc.edu}
 \affiliation{Laboratori de Càlcul Numeric (LaCàN), Universitat Politècnica de Catalunya-BarcelonaTech, Barcelona, Spain.}
\author{Tom Golde}
 \affiliation{Institute for Bioengineering of Catalonia (IBEC), The Barcelona Institute of Science and Technology (BIST), Barcelona, Spain.}
\author{Xavier Trepat}
 \affiliation{Institute for Bioengineering of Catalonia (IBEC), The Barcelona Institute of Science and Technology (BIST), Barcelona, Spain.}
 \affiliation{Facultat de Medicina, Universitat de Barcelona, Barcelona, Spain.}
 \affiliation{Institució Catalana de Recerca i Estudis Avançats (ICREA), Barcelona, Spain.}
 \affiliation{Centro de Investigación Biomédica en Red en Bioingeniería, Biomateriales y Nanomedicina (CIBER-BBN), Barcelona, Spain.}
\author{Marino Arroyo}%
\email[Email: ]{marino.arroyo@upc.edu}
 \affiliation{Laboratori de Càlcul Numeric (LaCàN), Universitat Politècnica de Catalunya-BarcelonaTech, Barcelona, Spain.}
 \affiliation{Institute for Bioengineering of Catalonia (IBEC), The Barcelona Institute of Science and Technology (BIST), Barcelona, Spain.}
 \affiliation{Centre Internacional de Mètodes Numèrics en Enginyeria (CIMNE), Barcelona, Spain.}

\begin{abstract}
Inspired by massive intermediate filament (IF) reorganization in superstretched epithelia, we examine computationally the principles controlling the mechanics of a set of entangled filaments whose ends slide on the cell boundary. We identify an entanglement metric and threshold beyond which random loose networks {respond non-affinely and nonlinearly to stretch by  self-organizing} into structurally optimal star-shaped configurations. A simple  model connecting cellular and filament strains links emergent mechanics to cell geometry, network topology, and filament mechanics. We identify a safety net mechanism in IF networks and provide a framework to harness entanglement in soft {fibrous} materials.

\end{abstract}
\maketitle

\setcounter{page}{1}

{Epithelial tissues are cohesive cellular sheets lining free surfaces in {multicellular eukaryotes}. They are involved in crucial physiological processes such as morphogenesis, protection,  secretion and absorption \cite{Bosveld2012,Alberts2014,He2014,Latorre2018}. Being biological barriers, they need to preserve integrity within active and challenging mechanical environments. Depending on the temporal scales and system, epithelial mechanics may depend on cellular rearrangements or on deformation of individual cells, which in turn depends on intracellular cytoskeletal networks that are mechanically integrated at the tissue scale through cell-cell junctions \cite{lazarides1980intermediate,Herrmann2007,harris2012characterizing,efimova2018branched}. These cytoskeletal networks are composite systems combining widely diverse biopolymers, which interact chemically, physically and through biological regulation  \cite{Huber2015,ndiaye2022intermediate,Chang2004,chang2014effects}.}

{Cytoskeletal actin filaments are short ($< 1$ \si{\micro\meter} \cite{chaudhuri2007reversible}), stiff both to stretch and bending \cite{janmey1991viscoelastic,broedersz2014modeling,Huber2015}, bind to a variety of specific crosslinkers including myosin motors, and turnover within minutes \cite{Huber2015}. Microtubules are long, stiff, dynamic, and also bind to specific motors. Conversely, intermediate filaments (IFs) organize into long (several \si{\micro\meter}  \cite{wagner2007softness}), bendable \cite{broedersz2014modeling,Huber2015} and highly stretchable bundles, with a highly nonlinear force-stretch behavior enabling extensions of up to $3-4.5$ fold extension \cite{kreplak2005exploring,kreplak2008tensile,qin2009hierarchical,qin2011flaw,block2017nonlinear,torres2019combined}. IF turnover is much slower \cite{Huber2015,Khalilgharibi2019}, in the order of hours, and unlike other cytoskeletal filaments, they lack stable and specific linkers, although unspecific cytolinkers such as plectin bind IFs to other IFs including nuclear lamins, to actin, or to microtubules  \cite{block2015physical}. IFs form supracellular networks thanks to adhesion complexes known as desmosomes \cite{Herrmann2007}. Together, these features support the view that IFs form a relatively passive network providing a ``safety belt'' against fast and large deformations \cite{wang2002mechanics,fudge2008intermediate,qin2009hierarchical,block2017nonlinear,golde2018glassy}, although how load is transferred from the tissue scale to individual IFs remains poorly understood.}

\begin{figure}[b]
\centering
\includegraphics[scale=0.29]{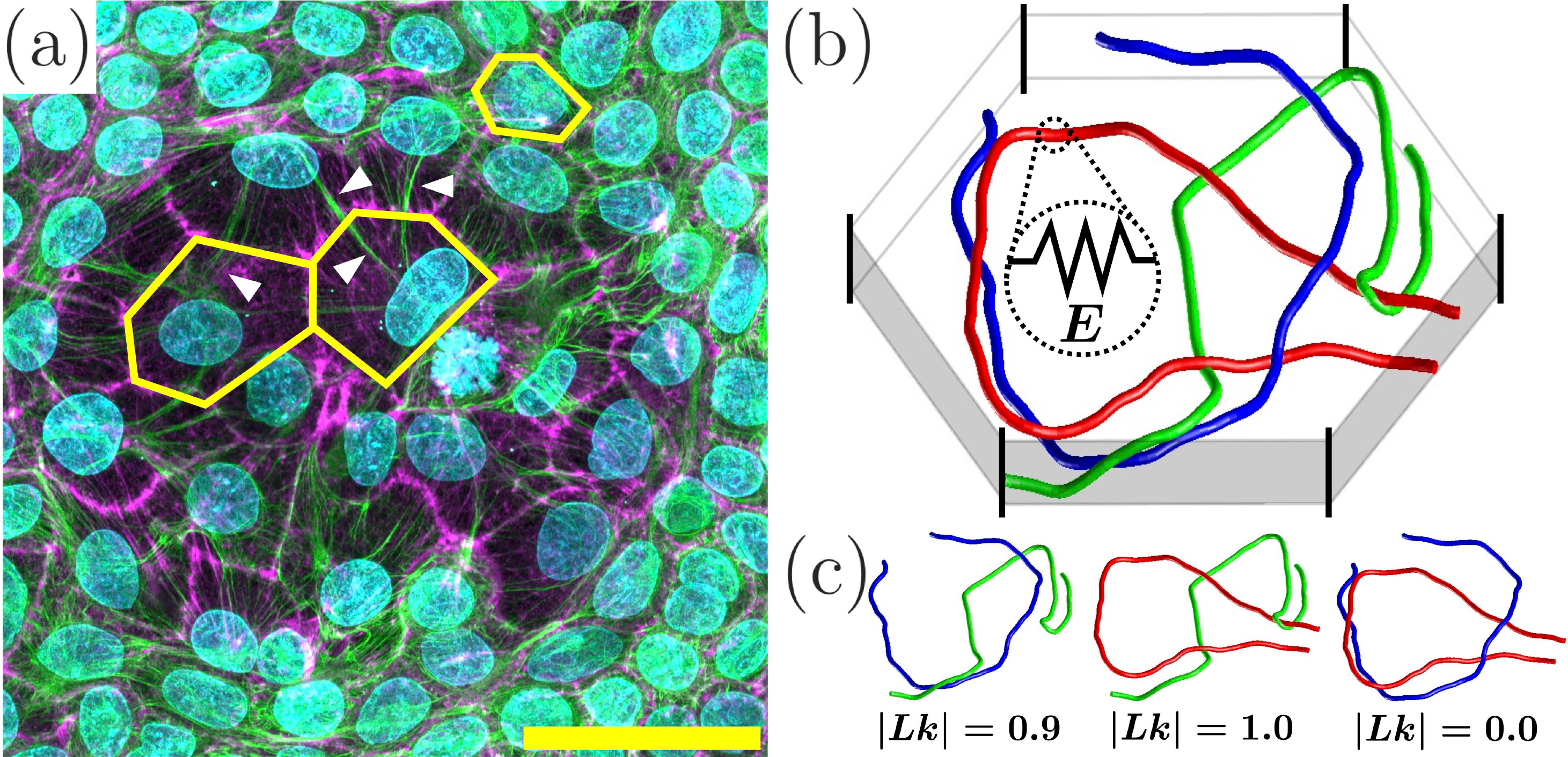}
\caption{\label{fig:Fig1} (a) Superstretched cells in an epithelial dome \cite{Latorre2018} with radial IF bundles (white arrowheads; green: keratin 18 IFs; magenta: actin; cyan: nuclei; scale bar: 40 \si{\micro\meter}), where yellow polygons mark selected cell outlines. (b) Sketch of a discrete network model of IFs in a cell with elastic, bendable filaments whose ends slide on the cell boundary. (c) Illustration of the pairwise Gaussian linking number for the filaments in (b).}
\end{figure}

{Recent stretching experiments on epithelial monolayers suggest synergistic interactions between the actin cytoskeleton, controlling epithelial mechanics at moderate stretches, and IFs,  providing load bearing under very large cell deformations \cite{harris2012characterizing,Latorre2018,Khalilgharibi2019,Duque2023.01.05.522736}. Under extreme cellular stretches and over long times, the IF network rearranges into a characteristic ``star-shaped'' structure where thick bundles radiate from a central tight tangle, Fig.~\ref{fig:Fig1}(a) \cite{Latorre2018}. Laser ablation shows that such IF bundles provide structural integrity to superstretched cells as the actin network becomes progressively diluted. Since IFs appear wavy and cortically arranged in unloaded cells, we wondered about the physical mechanisms underlying this slack-taut transition and the corresponding effects on network-scale mechanics.}

{To address these questions, we focused on the mechanics of IF networks alone. Although actin or nuclear lamins may limit the reconfigurations of the IF network by entanglements and crosslinks, the  separation of turnover time-scales mentioned earlier justifies a slow loading regime in which these constraints have time to relax but entanglements between IFs and their binding to desmosomes remain intact. Given the extreme extensibility of IFs \cite{lorenz2019lateral,kreplak2005exploring,kreplak2008tensile,qin2009hierarchical,qin2011flaw,block2017nonlinear,torres2019combined}, we ignore filament damage. We thus} idealize the IF cytoskeleton within a cell as a loose and entangled network of uncrosslinked, bendable, and extensible filaments, whose ends are attached to the lateral boundaries of a prismatic domain, Fig.~\ref{fig:Fig1}(b). {These idealized desmosomes prevent unraveling of the network by filament reptation. Since the IF network is corralled into cellular compartments, we regard} this cell as a {minimal mechanical unit} (yellow hexagons in Fig.~\ref{fig:Fig1}(a)). 

We modeled such networks using the cytoskeletal simulation suite \emph{cyto\textbf{sim}} \cite{nedelec2007collective,cytosimGitHub} for the Brownian dynamics of inextensible and bendable filaments, which we customized to model extensible filaments with general constitutive relations \cite{cytosimPensalfini}. We prepared computational models comprising $N_f$ cylindrical filaments of {persistence length $\ell_p$, reference length $\ell_0 \gg \ell_p$, and diameter $\phi \ll \ell_0$, according to the procedure described in Section \ref{sec:ModelGen} in \cite{supplement}. We initially considered linearly elastic filaments with modulus $E$. All filament points are confined inside the cell volume, interact sterically to avoid mutual crossing, and are subjected to a drag force with coefficient $\nu$,  Section \ref{sec:DefRate} in \cite{supplement}. We modeled the cell as a right regular prism whose base has $N_e$ edges, apothem length $a_0$, surface area $A_0$, and side length $s_0 = 2 a_0  \tan(\pi/N_e)$; the prism height is $h_0=a_0/4$. The model parameters and their rationale are provided in Table~\ref{supp_table} and Section \ref{sec:ModelParams} in \cite{supplement}, although our main results are rather insensitive to material parameters.}

To simulate the extreme equibiaxial stretching {reached by individual cells in pressurized lumens in vitro \cite{Latorre2018} and in developing embryos \cite{Deglincerti:2016aa}, we gradually increased the cell area, $A$, by $11$-fold at a slow strain rate, much smaller than 
the inverse intrinsic time constant of the system, $E/\nu$.}  During stretching, filament ends remain attached to the lateral boundaries such that the attachment locations may move laterally within a face but not to neighboring faces, in agreement with the notion that desmosomes can move laterally in adhered pairs of plasma membranes but cannot cross tricellular junctions. As we change $A$, and hence the areal strain, $\varepsilon_c = A/A_0 - 1$, we quantify the forces acting on the side walls. If $\bm{F}_i^{+}$ and $\bm{F}_i^{-}$ are the forces on the $i-$th filament ends, and $\hat{\bm{n}}_i^{+}$ and $\hat{\bm{n}}_i^{-}$ the normals to the walls that constrain those ends, the total force is $F_c = \sum_{i=1}^{N_f}{\left( \bm{F}_i^{+} \cdot \hat{\bm{n}}_i^{+} + \bm{F}_i^{-} \cdot \hat{\bm{n}}_i^{-} \right)}$, from which we define the nominal cellular tension, $T_c = F_c / N_e s_0$, and its dimensionless equivalent, $T_c^* = F_c \, a_0 / EA_f N_e s_0$. 

Lacking crosslinkers, entanglement is the only mechanism for our idealized networks to develop mechanical resistance. Thus, we established a system preparation protocol allowing us to control network entanglement by modifying the fraction of time during which filaments grow unconstrained or with their ends fixed to cell walls, Section \ref{sec:ModelGen} in \cite{supplement} and Movie S1. In agreement with our rationale, a loose and randomly organized network with default parameters and strong entanglement undergoes a dramatic spontaneous reorganization when stretched, Movie S2, leading to a central tight tangle from which filament bundles radiate perpendicularly to the lateral cell boundaries. The formation of such ``star-shaped'' configuration, reminiscent of IFs in superstretched epithelial cells, involves significant lateral motion of the attachment points and results in all filaments carrying load.
Conversely, an equivalent system with low entanglement develops less predictable and directed network reorganizations, where only a small fraction of filaments become taut under stretch, Movie S3. 

{The study of how entanglements restrict configurational entropy and hence the elastic properties of bulk polymeric materials has a long history \cite{Edwards_1967, deam1976theory, iwata1989new, everaers2004rheology}. Here instead, we sought to characterize the topological conditions for the self-organization of corralled entangled networks into star-shaped states under finite stretch. To define a predictive entanglement metric, we resorted to topological invariants, which mathematically characterize knots (embeddings of the unit circle in $\mathbb{R}^{3}$) and links (collection of knots) \cite{chmutov2012introduction,grishanov2012characterisation} and have been used to describe the topology of proteins and DNA \cite{ernst1995analysis,marko2015biophysics,niemyska2020gln}.  The pairwise Gaussian linking number, $Lk_{i,j}$, characterizes the number of times that a closed and oriented spatial curve, $\delta_i$, winds around another oriented curve, $\delta_j$, and can be computed as \cite{ricca2011gauss,chmutov2012introduction}
\begin{equation}\label{eq:GLNexact}
    \begin{split}
        & Lk_{i,j} =\\
            &\frac{1}{4\pi} \int_{0}^{2\pi} \int_{0}^{2\pi} \frac{\mathbf{r_i}(t_i) - \mathbf{r_j}(t_j)}{\left|\mathbf{r_i}(t_i) - \mathbf{r_j}(t_j)\right|^3} \cdot \left[\mathbf{r_i'}(t_i) \times \mathbf{r_j'}(t_j)\right] \, dt_i \, dt_j,
    \end{split}
\raisetag{3.5\normalbaselineskip}
\end{equation}
where $\mathbf{r_i}(t)$ and $\mathbf{r_j}(t)$ with $t \in [0,2\pi)$ are parameterizations of $\delta_i$ and $\delta_j$, and the prime denotes differentiation with respect to $t$. For links, $Lk_{i,j}$ is an integer and is invariant with respect to  deformations respecting mutual filament avoidance. For pairs of open curves with fixed ends, called \emph{tangles} \cite{chmutov2012introduction,grishanov2012characterisation}, $Lk_{i,j}$ is not a strict invariant, but it is still suitable to characterize pairwise linking \cite{marko2015biophysics}, Fig.~\ref{fig:Fig1}(c) and Section \ref{sec:LnkDsct} in \cite{supplement}. }

{To fully characterize topology in networks containing many filaments, one should resort to multi-body invariants beyond the pairwise Gaussian linking number. Classical work in polymer physics avoids this full enumeration, and instead simplifies the topological description using two-body invariants only \cite{Edwards_1967, deam1976theory, iwata1989new}. Accordingly, we considered the \emph{total pairwise Gaussian linking number} of the network, $Lks = \sum_{j>i} \left| Lk_{i,j} \right|$, previously adopted for textiles \cite{grishanov2009topological2}. However, neither this entanglement metric nor the \emph{average pairwise Gaussian linking number per number of filaments}, $Lks/N_f$,  predict whether a network is sufficiently entangled to self-organize into a star-shaped organization under stretch independently on the number of filaments $N_f$, Section \ref{sec:entanglement} in \cite{supplement}. Instead, we found that the \emph{average pairwise Gaussian linking number per number of filament pairs}, 
\begin{equation}\label{eq:avgGLN}
    \mathcal{E} = \frac{Lks}{N_p} = \frac{2}{N_f (N_f-1)} \sum_{j>i} \left| Lk_{i,j} \right| ,
\end{equation}
systematically discerns between the two network behaviors illustrated in Movies S2 and S3, Section \ref{sec:entanglement} in \cite{supplement}. While $\mathcal{E}$ is not a strict topological invariant, we verified that it is essentially independent of network deformation, Section \ref{sec:entanglementVsStrain} in \cite{supplement}.}

\begin{figure}
\centering
\includegraphics[scale=0.42]{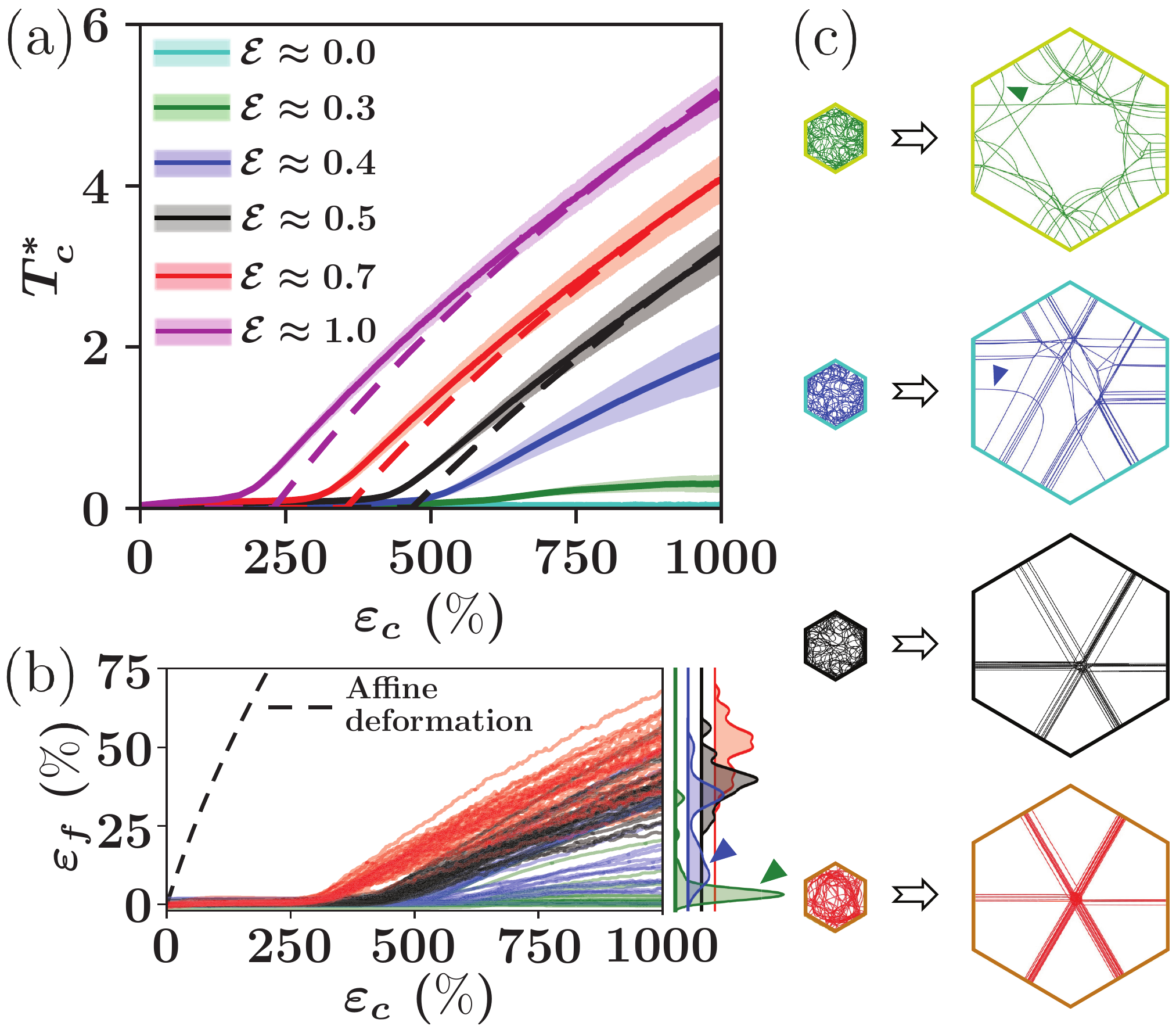}
\caption{\label{fig:Fig2} Influence of filament entanglement on cellular mechanical response. (a) Relation between normalized cellular tension, $T_c^*$, and  cell areal strain, $\varepsilon_c$ (solid lines and shadings: mean $\pm$ standard deviation of $8$ model realizations; dashed lines: 1D analytical model with $\gamma = 1.0$, $2.2$, $2.9$ for $\mathcal{E}\approx0.5$, $0.7$, $1.0$). {(b) Evolution of individual filament strain $\varepsilon_f$ as a function of cell areal strain for representative realizations of different levels of entanglement and corresponding distributions of filament strain at $\varepsilon_c = 1000\%$ (right shaded curves). The dashed line is the prediction $\varepsilon_f = \sqrt{\varepsilon_c+1}-1$ under the assumption of affinity.} (c) Representative network reorganizations for $\varepsilon_c = 1000\%$.}
\end{figure}

We then systematically examined the role of entanglement on the network mechanics by considering filament ensembles with varying degree of entanglement $\mathcal{E}$ and otherwise identical model parameters. For insufficiently entangled systems ($\mathcal{E} \lesssim 0.3$), the networks do not exhibit coherent reorganization, Movies S3 and S4, and concomitantly, the buildup of tension is insignificant (cyan and green curves in Fig.~\ref{fig:Fig2}(a)) in line with previous findings on non-woven textiles \cite{Negi}. Since modest levels of $\mathcal{E}$ correspond to limited mutual winding, IFs not directly bridging opposite sides can accommodate cellular deformations without elongating (green curves in Fig.~\ref{fig:Fig2}(b)). By remaining slack (green arrowheads in Fig.~\ref{fig:Fig2}(b,c)), these filaments do not contribute to the emergent mechanical response.

For $\mathcal{E} \approx 0.4$, the networks develop several tight tangles connecting taut filaments, Movie S5. This topological reorganization enables sustained cellular stiffening, blue curve in Fig.~\ref{fig:Fig2}(a). However, the filament strain distribution is extremely broad, indicating that some are strongly elongated while others remain slack, blue arrowheads in Fig.~\ref{fig:Fig2}(b,c).

For $\mathcal{E} \gtrsim 0.5$, the networks robustly reorganize into star-shaped configurations, Movies S2 and S6, mobilizing all filaments with similar elongation, Fig.~\ref{fig:Fig2}(b), and stiffening beyond an activation strain $\varepsilon_c^A$, Fig.~\ref{fig:Fig2}(a). We infer that entanglement enables self-organization of the networks into structurally optimal filament arrangements, akin to IF reorganization in superstretched epithelial cells \cite{Latorre2018}, Fig.~\ref{fig:Fig1}(a). The transition of system behavior beyond a critical degree of entanglement can be interpreted as a {topological threshold for mechanical self-organization. Importantly, thanks to their strong non-affinity, these structurally-optimal networks offer mechanical resistance to extreme cell deformations with moderate individual filament strains, defined as $\varepsilon_f = \ell/\ell_0-1$ with $\ell$ the current filament length,  Fig.~\ref{fig:Fig2}(b). For instance, for $\mathcal{E} \approx 0.5$, cell areal strains of $\varepsilon_c =1000$\% are accommodated by filament strains of about $\varepsilon_f = 40$\%, much lower than the filaments strains around 230\% of an equally stretched affine network}.

To understand the parameters controlling $\varepsilon_c^A$ and the subsequent tension-strain relation, we developed an analytical model assuming {network entanglement above the topological threshold, Section \ref{sec:AnalyticalModel} in \cite{supplement}}. Considering the star-shaped geometry of the stretched network, and accounting for the filament length stored in the central tight tangle, this model links cell- and filament-scale deformations to estimate the cellular activation strain,
\begin{equation}\label{eq:activStrain}
    \varepsilon_c^A \approx \frac{1}{4 a_0^2} \left[ \ell_0 - \frac{\pi}{4} \phi \, \gamma(\mathcal{E}) \, \mathcal{E} \left(N_f - 1\right) \right]^2 - 1,
\end{equation}
where $\gamma(\mathcal{E})$ is a phenomenological scalar quantifying the average complexity of individual windings within the central tight tangle. $\gamma = 1$ models a situation in which windings involve filament pairs. With increasing entanglement, we expect windings to involve more filaments, and hence require more length, leading to larger $\gamma$. Neglecting filament bending, our model also provides an approximate expression for the dimensionless nominal cellular tension: 
\begin{equation}\label{eq:cellForce_an}
    T_c^* \approx \frac{2 \, N_f \, a_0}{\ell_0 \, N_e \, \tan(\pi/N_e)} \left\langle \sqrt{\varepsilon_c+1} - \sqrt{\varepsilon_c^A+1} \; \right\rangle,
\end{equation}
where {the angle brackets of a real number $a$ are defined by} $\langle a \rangle$ = 0 if $a<0$ and $\langle a \rangle = a$ otherwise. Thus, the only fitting parameter is $\gamma(\mathcal{E})$, which should be close to 1 for networks barely above the topological threshold  and increase with $\mathcal{E}$, Section \ref{sec:AnalyticalModel} in \cite{supplement}.

\begin{figure*}
\centering
\includegraphics[scale=0.64]{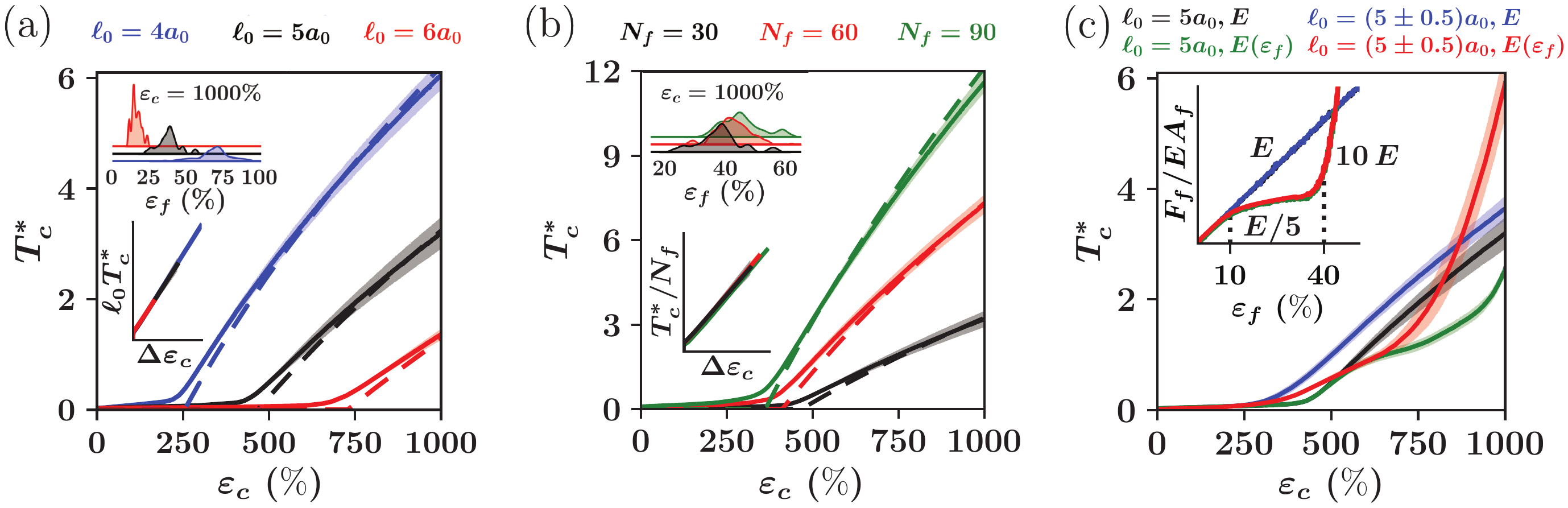}
\caption{\label{fig:Fig3} Cellular mechanical response (solid lines and shadings: mean $\pm$ standard deviation of $8$ model realizations; dashed lines: 1D analytical model with $\gamma=1${; insets: $\Delta\varepsilon_c = \langle \sqrt{\varepsilon_c + 1} - \sqrt{\varepsilon_c^A + 1} \, \rangle $}) and filament strain distribution at $\varepsilon_c=1000\%$ when varying $\ell_0$ (a), $N_f$ (b), the distribution of filament lengths $\ell_0$ and the filament strain-force relation (c). In all cases, $\mathcal{E}\approx 0.5$.}
\end{figure*}

We found a nearly quantitative match between the analytical model with $\gamma= 1$ and simulations at the threshold $\mathcal{E}\approx 0.5$. For higher entanglement, we found very good agreement by increasing $\gamma$ to 2.2 and 2.9 for $\mathcal{E}\approx 0.7$ and $\mathcal{E}\approx 1.0$, Fig.~\ref{fig:Fig2}(a), consistent with the idea that networks with larger $\mathcal{E}$ involve windings of increasing complexity. In agreement with our analytical model, the mechanical response of the system {above the topological threshold},  particularly the emergent stiffness $\partial T_c^*/\partial \varepsilon_c$, is essentially independent of entanglement except for the shift in $\varepsilon_c^A$. Accordingly, we considered a default entanglement $\mathcal{E} \approx 0.5$ in subsequent simulations and set $\gamma = 1$ for the theoretical fits.

According to our theory, as more filament length is stored in the central tight tangle, less length is available for the bundles to bridge cell boundaries. As a result, {increasing entanglement should not only increase $\varepsilon_c^A$ but also individual filament strain $\varepsilon_f$ for a given cellular strain, in agreement with our simulations (black and red curves in Fig.~\ref{fig:Fig2}(b))}. Additional simulations show that the mechanical response and network mechanisms described here are not modified by thermal vibrations, Section \ref{sec:Temperature} in \cite{supplement}, or changes in filament bending rigidity, Section \ref{sec:Bending} in \cite{supplement}, and that the emergent stiffness scales proportionally to the filament elastic modulus, Section \ref{sec:Emodulus} in \cite{supplement}.

To further test our theory, we examined the role of filament length, which according to Eqs.~(\ref{eq:activStrain}, \ref{eq:cellForce_an}), should modify the activation strain, $\varepsilon_c^A$, and the emergent tension, $T_c^*$. In agreement with the analytical predictions, simulations with shorter/longer filaments lead to smaller/larger activation strains and stiffer/softer networks, Fig.~\ref{fig:Fig3}(a), with a marked downward shift in  filament strain distributions for longer filaments as more filament length is available to accommodate cellular strain. Per Eq.~(\ref{eq:cellForce_an}), the tension-strain curves of networks with different filament lengths collapse when representing  $\ell_0 T_c^*$ as a function of $\sqrt{\varepsilon_c + 1} - \sqrt{\varepsilon_c^A + 1}$, Fig.~\ref{fig:Fig3}(a)-inset, reflecting the increased compliance of longer filaments. Varying the number of filaments, $N_f$, linearly affects the slope of the cellular response, Fig.~\ref{fig:Fig3}(b), in agreement with Eq.~(\ref{eq:cellForce_an}).
Instead, the number of filaments mildly impacts the activation strain,  Fig.~\ref{fig:Fig3}(b).
Thus, filament loading in our entangled networks is determined primarily by $\mathcal{E}$ and $\ell_0$, and only weakly by $N_f$, whereas emergent tension and stiffness are directly controlled by $N_f$ and $\ell_0$. 

Since IFs exhibit a highly nonlinear force-stretch relation, we then wondered whether the filament constitutive behavior affected the slack-taut transition and the emergent mechanics. We considered filaments that soften to $E/5$ for $\varepsilon_f$ in the range between 10\% and 40\% and eventually re-stiffen to reach $10E$, mimicking their typical superelastic response \cite{kreplak2005exploring,kreplak2008tensile,qin2009hierarchical,qin2011flaw,block2017nonlinear,torres2019combined}. The emergent stiffness of the taut network mirrors the individual IF constitutive relations, Fig.~\ref{fig:Fig3}(c), consistent with the narrow filament strain distributions observed for $\mathcal{E}\gtrsim0.5$, Fig.~\ref{fig:Fig2}{(b)}. To examine the influence of filament length variability, we sampled $\ell_0$ from a normal distribution with mean $5a_0$ and standard deviation $0.5 a_0$. For linearly elastic filaments, this reduces the activation strain, as shorter filaments are mobilized earlier. For nonlinear filaments, the plateau in the filament response is lost at the cellular scale, as the emergent behavior now results from convolving the nonlinear constitutive laws of unequally strained filaments. Cell-scale stiffening is also reached earlier when including shorter filaments. However, the slack-taut transition corresponding to the formation of radial IF bundles remains unchanged, Section \ref{sec:TransitionConstitutive} in \cite{supplement}. Thus, while the shape of the emergent mechanical response past the activation strain depends on the constitutive behavior, the number, and the length distribution of the filaments, their non-affine self-organization into a star-shaped configuration is solely determined by network entanglement. 

\begin{figure}[t]
\centering
\includegraphics[scale=0.29]{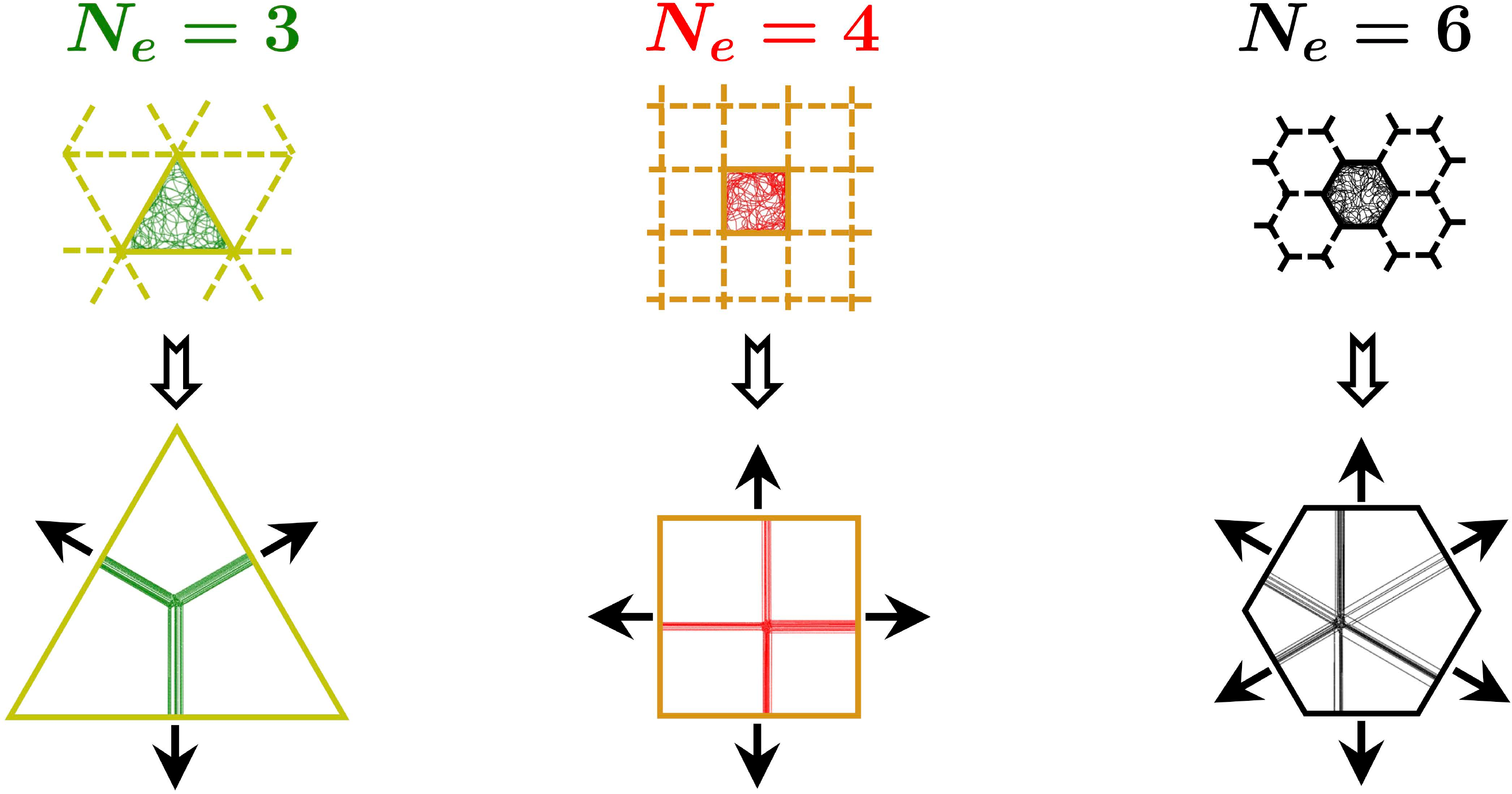}
\caption{\label{fig:Fig4} Network reorganization for different shapes of the enclosing cell, corresponding to regular tessellations of the plane.}
\end{figure}

Following this rationale, the slack-taut transition should also be preserved when varying the cell shape. To test this, we prepared networks with default parameters and enclosed them in cells with the same $a_0$ but different $N_e$. According to the three regular tessellations of the plane \cite{grunbaum1977tilings}, we compared triangular, square, and hexagonal cells ($N_e = 3,4,6$). Remarkably, {the non-affine and nonlinear mechanical response above the topological threshold is independent of cell shape and well described by Eqs.~(\ref{eq:activStrain}, \ref{eq:cellForce_an}), Fig.~\ref{fig:Fig4} and Section \ref{sec:TransitionOtherShapes} in \cite{supplement}.}

In summary, inspired by the phenomenology of IF networks in epithelial monolayers under stretch \cite{Latorre2018}, we have studied the physical principles supporting the {nonlinear and non-affine mechanical response} of an ensemble of entangled extensible filaments confined to a cell with laterally moving boundary attachments. We identify  a metric of entanglement, $\mathcal{E}$, which robustly predicts a threshold {for mechanical activation of all filaments,} $\mathcal{E} \gtrsim 0.5$, leading to self-organization of random filament networks into structurally optimal configurations beyond an activation strain. The occurrence of the transition is purely topological, whereas the emergent mechanics depend on length, number, and constitutive response of the filaments, enabling independent control of activation strain and stiffness.

Our work suggests that, through entanglement and self-organization, IF networks provide a  ``safety net'' for cells against extreme strains. Being rooted in network topology, this emergent material property is complementary to the role of IFs as a ``safety belt'' against fast strain rates, which hinges on rate-dependent mechanics of filaments and bundles \cite{block2017nonlinear}. Beyond the biological context, network entanglement has been leveraged to enhance the mechanical properties of hydrogels \cite{kim2021fracture,liu2021tough} and is at the core of textile materials \cite{jeon2014situ,moyo2017micromechanics,liu2019multiscale}. Here, we identify  \emph{corralled entanglement} as a scale-free principle for extremely deformable bioinspired materials whose organization lies between random networks and woven materials. By relying on self-organization, this principle is devoid of the synthesis challenges of weaving or knitting at a molecular scale \cite{Zhang:2022aa}.

\let\oldaddcontentsline\addcontentsline
\renewcommand{\addcontentsline}[3]{}
\begin{acknowledgments}
\vspace{0.1cm}
\noindent \textbf{Acknowledgments.} We  acknowledge the computer resources at Caléndula (SCAYLE) and the technical support provided by Barcelona Supercomputing Center (RES-IM-2021-2-0017 \& RES-IM-2021-2-0028). This work was supported through funding from the Spanish Ministry of Science and Innovation \& NextGenerationEU/PRTR (PCI2021-122049-2B), the EU Research Council (CoG-681434), the EU Commission (H2020-FETPROACT-01-2016-731957), and the German Research Foundation (DFG GO3403/1-1).
\end{acknowledgments}
\let\addcontentsline\oldaddcontentsline

\let\oldaddcontentsline\addcontentsline
\renewcommand{\addcontentsline}[3]{}
\let\addcontentsline\oldaddcontentsline


\onecolumngrid

\newpage

\thispagestyle{empty}

\begin{center}
  \textbf{\large Supplemental Material: \\ Non-affine mechanics of entangled networks inspired by intermediate filaments}\\[.2cm]
  Marco Pensalfini,$^{1}$ Tom Golde,$^{2}$ Xavier Trepat,$^{2,3,4,5}$ and Marino Arroyo$^{1,2,6}$\\[.1cm]
  {
  \itshape ${}^1$ Laboratori de Càlcul Numeric (LaCàN), Universitat Politècnica de Catalunya-BarcelonaTech, Barcelona, Spain.\\
  ${}^2$ Institute for Bioengineering of Catalonia (IBEC), \\The Barcelona Institute of Science and Technology (BIST), Barcelona, Spain.\\
  ${}^3$ Facultat de Medicina, Universitat de Barcelona, Barcelona, Spain.\\
  ${}^4$ Institució Catalana de Recerca i Estudis Avançats (ICREA), Barcelona, Spain.\\
  ${}^5$ Centro de Investigación Biomédica en Red en Bioingeniería, \\Biomateriales y Nanomedicina (CIBER-BBN), Barcelona, Spain.\\
  ${}^6$ Centre Internacional de Mètodes Numèrics en Enginyeria (CIMNE), Barcelona, Spain.\\
  }
\end{center}

\setcounter{equation}{0}
\setcounter{secnumdepth}{4}
\setcounter{page}{1}
\renewcommand{\theequation}{S\arabic{equation}}
\renewcommand{\thesection}{S\arabic{section}}  
\renewcommand{\thepage}{S\arabic{page}}
\renewcommand{\thefigure}{\thesection}
\renewcommand{\thetable}{\thesection}

\onecolumngrid

\vspace{1cm}

\tableofcontents

\newpage

\section{\label{sec:ModelGen} Generation of entangled filament network models}

\begin{figure*}[h]
\centering
\includegraphics[scale=0.6]{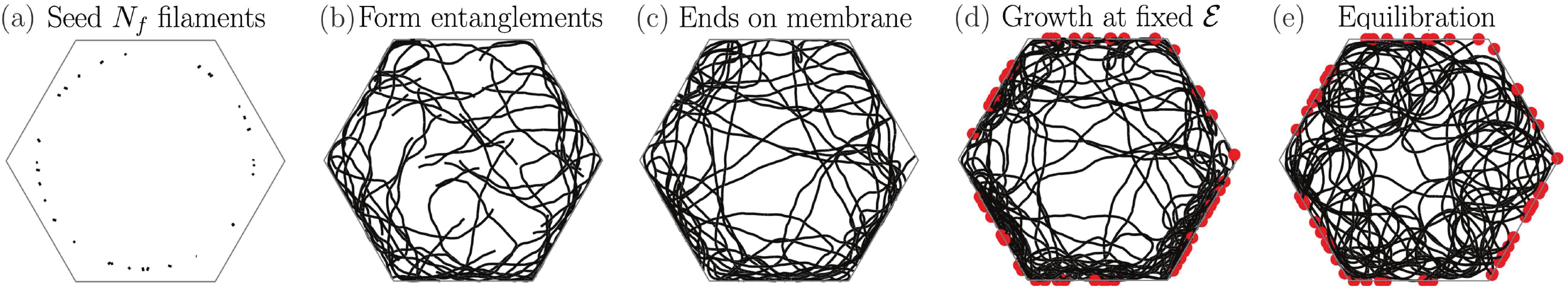}
\caption{\label{fig:FigS1} Model generation at room temperature. $N_f$ filaments are seeded within the confining space (a) and let grow to form entanglements (b). Upon reaching the prescribed {value of} $\mathcal{E}$, a confining potential brings the filament ends on the {lateral} cell walls (c). Then, $\ell_0$ is  adjusted to reach the desired value by letting filaments grow while their ends are fixed, so that $\mathcal{E}$ cannot change significantly (d). A long equilibration phase (e) ensures that any pretension possibly associated with model generation is removed prior to stretching for mechanical characterization.}
\end{figure*}

To establish computational models of entangled filament networks, we leveraged \emph{cyto\textbf{sim}}'s ability to simulate the growth of inextensible filaments in a confined cellular space{. We model the cell} as a right regular prism {whose} base has $N_e$ edges {and apothem length $a_0$ in the reference configuration; the cell height is set to $h_0 = a_0/4$.}
We start by seeding $N_f$ {cylindrical} filaments of initial length {$\ell_0^i \ll \ell_0$} and bending rigidity {$\kappa = \ell_pk_BT$, $\ell_p$ being the persistence length of one IF bundle and $k_BT$ the thermal energy of the system,} on the surface of a cylinder of radius {$R < a_0$, which is} coaxial with the prism {and has height $h_0$}, Fig.~\ref{fig:FigS1}(a).
The filament diameter, {$\phi \ll a_0$}, is represented by enforcing repulsive steric interactions through a harmonic potential of stiffness {$k_s \gg E\phi^2/\ell_0$}, resulting in frictionless contacts among filaments.
All filament points are confined inside the cell volume by a harmonic potential whose stiffness is set to {$k_c \gg E\phi^2/\ell_0$} for points outside the cell volume, {with $k_c < k_s$,} and to $0$ for points located inside.
{The numerical values of all model parameters are provided in Table~\ref{supp_table}.}

To form entanglements, we let filaments grow at room temperature ($k_BT = 0.0042$ \si{\pico\newton \micro\meter}) in an environment of effective viscosity {$\nu$}, Fig.~\ref{fig:FigS1}(b), resulting in a random-walk-like stochastic process {during which we track $\mathcal{E}$}. Importantly, filament ends are not constrained to lie  on the side walls of the cell at this stage.
Slightly before reaching the prescribed {level of entanglement}, we activate a confining potential, also of stiffness $k_c$, bringing each filament end to the {nearest lateral} face of the enclosing prism, Fig.~\ref{fig:FigS1}(c).
At this point, we adjust the length of the filaments  while holding their ends with very stiff springs {($k \gg k_s$)}, red dots in Fig.~\ref{fig:FigS1}(d), such that entanglement cannot change significantly. We then equilibrate the entire system at room temperature to eliminate any pretension that might arise during model generation, Fig.~\ref{fig:FigS1}(e). Since the described steps involve stochastic events, we consider $8$ model realizations for each parameter set. We treat the filaments as inextensible ($E = \infty$) during model generation. 

\clearpage

\newpage

{\section{\label{sec:ModelParams} Model parameters for network generation and stretching simulations}}

{To generate our networks  and subsequently simulate stretching, we need to assign numerical values to the model parameters relating to the physical and geometrical characteristics of the IF bundles, of the enclosing cell and its cytosolic environment, as well as to network topology. All adopted values are indicated in Table~\ref{supp_table}. Here,  we provide details on the rationale underlying our parameter choices.}

{For the cytosolic environment, we select an ambient viscosity $\nu = 1$ \si{\pico\newton \, \second / \micro\meter\squared}, in line with previous similar modeling studies \cite{descovich2018cross, cortes2020bond}, and perform model generation at room temperature ($k_BT = 0.0042$ \si{\pico\newton\micro\meter}). However, network stretching is simulated athermally ($k_BT \approx 0$ \si{\pico\newton\micro\meter}), which we confirm to be an acceptable approximation in Section \ref{sec:Temperature}.
Inspired by typical cell packings in epithelial monolayers, we model the cell enclosing IF bundles as an hexagonal prism ($N_e = 6$) in most of our calculations. The generality of the phenomenology that we identify with respect to the enclosing cell shape is confirmed in Fig.~\ref{fig:Fig4} and in Section \ref{sec:TransitionOtherShapes}, where we simulate networks enclosed within domains with a square or triangular base. Lastly, we select an apothem length $a_0 = 5$ \si{\micro\meter} for the cell base in the reference configuration, in line with previous modeling work \cite{Latorre2018}.}

{The default number of filaments included in the networks is set arbitrarily to $N_f = 30$, and is varied to show that this parameter does not affect the key qualitative features of the identified phenomenology, Fig.~\ref{fig:Fig3}(b), provided that the correct metric of entanglement is used, Section \ref{sec:entanglement}. Similarly, the default reference length of the filaments is arbitrarily set to $\ell_0 = 5a_0 = 25$ \si{\micro\meter}, and is varied to show the robustness of the nonlinear and non-affine network behavior, Fig.~\ref{fig:Fig3}(a). Such reference length is gradually reached in model generation simulations, starting from an initial value $\ell_0^i = 0.015$ \si{\micro\meter}; this value was selected to be much smaller than the default value $\ell_0$.}

{In our model, the physical properties of the IF bundles depend on fiber diameter $\phi$, Young's modulus $E$, and persistence length $\ell_p$. For the diameter, the literature reports a range of $40-130$ \si{\nano\meter} \textit{in vivo} \cite{nolting2014mechanics}. Accordingly, we took $\phi = 100$ \si{\nano\meter} for our simulations. For the Young's modulus used in cell stretching simulations, single filaments have values ranging from $6-300$ \si{\pascal} \cite{yoon2019keratin} to $1-10$ \si{\mega\pascal} \cite{fudge2003mechanical, stamenovic2011stress}. Since the larger values were obtained from experiments on hagfish threads, which might not be representative of IF mechanics within cells, we decided to rely on the lower values and reasoned that IF bundles should be stiffer than individual filaments. We took $E = 13$ \si{\kilo\pascal} as the default value for our simulations. Importantly, even when varying $E$ in a range spanning three orders of magnitude, Section \ref{sec:Emodulus}, or when making it dependent on the filament strain, Fig.~\ref{fig:Fig3}(c), we observe no fundamental difference in the nonlinear and non-affine emergent network mechanics. Lastly, the persistence length of IF bundles is hardly discussed in the literature, with a single study estimating indirectly a rather large value of 1 \si{\milli\meter} \cite{nolting2014mechanics}. Here, to select $\ell_p$, we started from typical values for single filaments ($0.2-1$ \si{\micro\meter} \cite{broedersz2014modeling}), which we increased by one order of magnitude to obtain the default parameter used in our simulations, $\ell_p = 5$ \si{\micro\meter}. The persistence length is related to the bending rigidity, used as an input parameter in \emph{cyto\textbf{sim}}, by $\kappa = \ell_p k_B T$, which we fixed at its room temperature value both in thermal and athermal simulations.}

{We note that, for a homogeneous beam of circular cross section, $E$, $\phi$, and $\kappa$ are related by $\kappa = \pi E \phi^4 / 64$. IF bundles are not homogeneous and may exhibit partial sliding \cite{nolting2014mechanics}, and hence these three material parameters may be chosen independently. Yet, with our default parameters justified above and shown in Table~\ref{supp_table}, we find $\pi E \phi^4 / 64 \approx 0.064$ \si{\pico\newton\micro\meter\squared}. This value is not too far from $\kappa = 0.021$ \si{\pico\newton\micro\meter\squared}. In any case, varying $\kappa$ in a range spanning four orders of magnitude, Section \ref{sec:Bending}, we observe no significant influence on the emergent network mechanics or self-organization.}

{Finally, in stretching simulations, we subject cells to an areal strain $\varepsilon_c = 10$, defined according to the experiments in \cite{Latorre2018}. To ensure quasi-static cell loading, see Section \ref{sec:DefRate}, the default strain rate, $\dot{\varepsilon}_c = 0.025$ \si{\per\second}, is selected to be much smaller than the inverse intrinsic time constant of the system, $E/\nu \geq 1300$ \si{\per\second}. }

\begin{table}[hb]
	\caption{{Model parameters for network generation and stretching simulations.}} 	
	\label{supp_table}
	\begin{center}
    		\begin{tabular}{ l | c | l | l}
			\textbf{Physical parameter}		& 	\textbf{Symbol}		& 	\textbf{Default value}										&	\textbf{Rationale}	\\ \hline
			Cell areal strain	 (max. value)	&   $\varepsilon_c$			&      $10$													&	Maximum value from Ref. \cite{Latorre2018}.	 \\ \hline
			Cell areal strain rate				&  $\dot{\varepsilon}_c$    &	$0.025$ \si{\per\second} 									&	\makecell[l]{Assumed ensuring that $\dot{\varepsilon}_c \ll E/\nu$. \\ Explored range $0.0125-2.5$ \si{\per\second} in Section \ref{sec:DefRate}.}	\\ \hline
			Cell reference apothem length		&	$a_0$			&	$5$ \si{\micro\meter}										&	Same order of magnitude as in Ref. \cite{Latorre2018}. \\ \hline
			Cell height					&	$h_0$			&      $1.25$ \si{\micro\meter}									&	Assumed $= a_0/4$.					\\ \hline
			Effective ambient viscosity		&	$\nu$			&  $1$ \si{\pico\newton \, \second / \micro\meter\squared}				&	Same value as in Refs. \cite{descovich2018cross, cortes2020bond}. 	\\ \hline
			IF bundle bending rigidity			&	$\kappa$			&	$0.021$ \si{\pico\newton \micro\meter\squared} 				&	\makecell[l]{$ = \ell_p k_B T$, \textit{cf.} Ref. \cite{broedersz2014modeling}. \\ Explored range $0.00021-2.1$ \si{\pico\newton \micro\meter\squared} in Section \ref{sec:Bending}.} \\ \hline
			IF bundle diameter				&	$\phi$			&      $100$ \si{\nano\meter}									&	Assumed within range measured in Ref. \cite{nolting2014mechanics}. \\ \hline
			IF bundle elastic modulus			&	$E$				& \makecell[l]{Generation: $\infty$ \\ Stretching: $13$ \si{\kilo\pascal} }	&	\makecell[l]{Assumed within broad ranges from Refs. \cite{fudge2003mechanical,stamenovic2011stress,yoon2019keratin}. \\ Explored range $1.3-1300$ \si{\kilo\pascal} in Section \ref{sec:Emodulus}. \\ Considered $E = E \left(\varepsilon_f\right)$ in Fig.~\ref{fig:Fig3}(c).}  \\ \hline
			IF bundle reference length			&	$\ell_0$			&	$5a_0$ 												&	\makecell[l]{Assumed ensuring that $\ell_0 > \ell_p$. \\ Explored range $4a_0-6a_0$ in Fig.~\ref{fig:Fig3}(a). \\ Random length distribution in Fig.~\ref{fig:Fig3}(c).}	 \\ \hline
			Stiffness of confining potential  		&	$k_c$			&      $10^3$ \si{\pico\newton / \micro\meter}						&	Assumed $= 50 EA_f / a_0$, with $A_f = \pi \phi^2 /4$.		 	\\ \hline
			Stiffness of springs holding IFs	        &	$k$				&      $10^6$ \si{\pico\newton / \micro\meter}						&	Assumed $= 1000 k_c$.				\\ \hline
			Stiffness of steric potential			&	$k_s$			&      $10^4$ \si{\pico\newton / \micro\meter}						&	Assumed $= 500 EA_f / a_0$, with $A_f = \pi \phi^2 /4$.			\\ \hline
			Thermal energy					&	$k_BT$			&	\makecell[l]{Generation: $0.0042$ \si{\pico\newton \micro\meter} \\ Stretching: athermal}	&      \makecell[l]{The influence of temperature on stretching \\ simulations is minor, \textit{cf.} Section \ref{sec:Temperature}.}		\\ \hline
		\end{tabular}
	\end{center}
\end{table}

\clearpage

\section{\label{sec:LnkDsct} Quantification of the pairwise linking number for closed and open curves}

\setcounter{figure}{2}

To evaluate entanglement in our computational models, where filaments are represented by sequences of segments, we approximate Eq.~(\ref{eq:GLNexact}) in the main text by \cite{niemyska2020gln}
\begin{equation}\label{eq:GLNapprox}
        Lk_{i,j} \approx \frac{1}{4 \pi} \sum_{A=1}^{N_i-1} \sum_{B=1}^{N_j-1} \frac{\mathbf{R_A^{(i)}} - \mathbf{R_B^{(j)}}}{\left|\mathbf{R_A^{(i)}} - \mathbf{R_B^{(j)}}\right|^3} \cdot [d\mathbf{R_A^{(i)}} \times d\mathbf{R_B^{(j)}}],
\end{equation}
where $\delta_i$ and $\delta_j$ have been discretized using $N_i$ and $N_j$ points that define  the piecewise straight segments $d\mathbf{R_A^{(i)}}$ and $d\mathbf{R_B^{(j)}}$ with midpoint locations $\mathbf{R_A^{(i)}}$ and $\mathbf{R_B^{(j)}}$.

\begin{figure*}[h]
\centering
\includegraphics[scale=0.67]{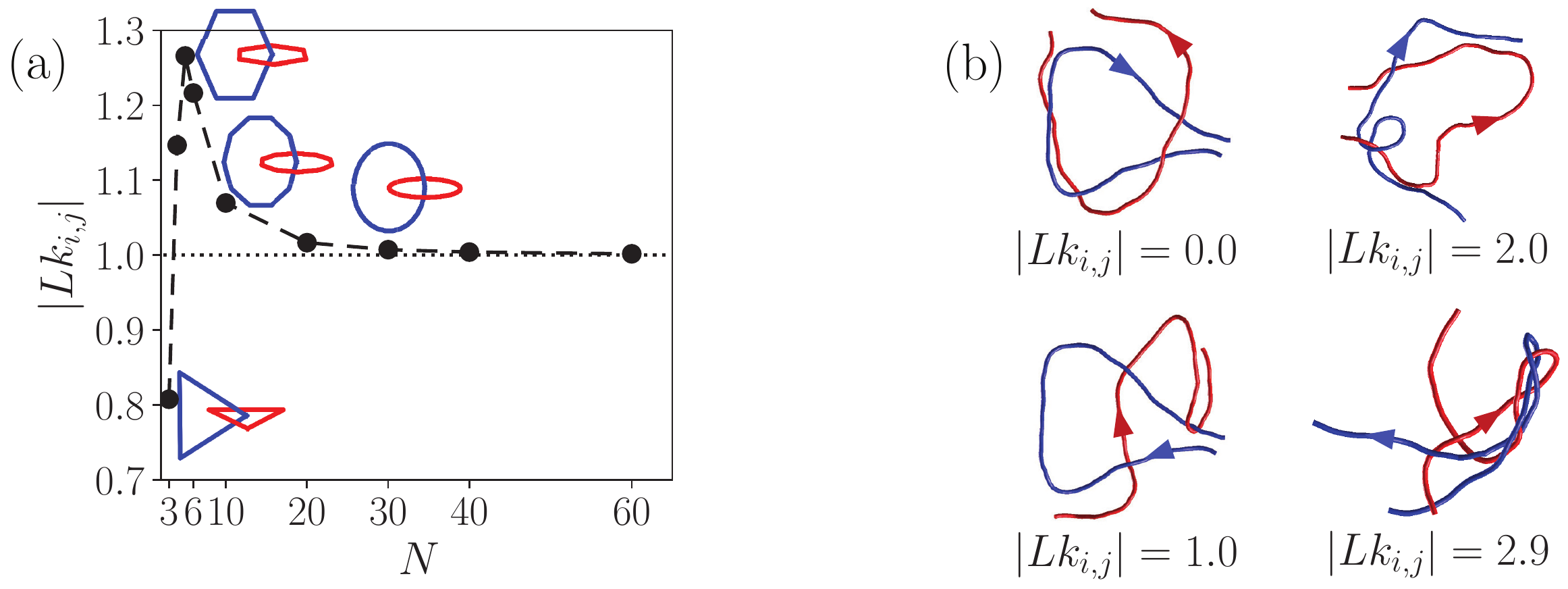}
\caption{\label{fig:FigS2} Quantification of the pairwise linking number for closed and open curves. For a pair of closed curves forming a Hopf link and discretized by a finite number of segments, $Lk_{i,j}$ converges to its exact value as the discretization is refined (a). For pairs of open curves, $Lk_{i,j}$ can be used to identify mutual winding despite not being a strict  topological invariant (b).}
\end{figure*}

To address the influence of such discretization on $Lk_{i,j}$, we consider a link with known linking number and apply Eq.~(\ref{eq:GLNapprox}) with varying $N_i$ and $N_j$, assuming for simplicity that $N_i = N_j = N$. For a Hopf link, \textit{i.e.} a set of two closed curves interlinked in the simplest possible way, the linking number is $1$. As shown in Fig.~\ref{fig:FigS2}(a), it is sufficient to take $N\geq10$ to keep the error induced by discretization below $10\%$. This confirms that a discrete version of Eq.~(\ref{eq:GLNexact}) in the main text can be used to provide a quite accurate estimate of $Lk_{i,j}$ for closed curves. However, the filaments that we model are open sequences of segments. Applying Eq.~(\ref{eq:GLNapprox}) to pairs of filaments akin to those in our network models shows that $Lk_{i,j}$ provides reasonable estimates of the degree of mutual winding between filaments, Fig.~\ref{fig:FigS2}(b).

\clearpage

\section{\label{sec:entanglement} Alternative network entanglement metrics and their link to mechanics}
To substantiate the use of the system-wide entanglement metric adopted in the main text, Eq.~(\ref{eq:avgGLN}), we compare alternative definitions and discuss their suitability to capture the characteristic self-organization of our filament ensembles. Specifically, we aim to show that $\mathcal{E}$ can correctly identify networks reorganizing into the characteristic star-shaped configuration, whereas  $Lks$ and $Lks/N_f$ cannot. We recall below the definitions of these entanglement metrics:
\begin{itemize}
\item  $Lks$ is the \emph{total} pairwise Gaussian linking number, approximating the total number of windings in the network if self-winding is neglected,
    \begin{equation}
        Lks = \sum_{j>i} \left| Lk_{i,j} \right| \approx N_w;
    \end{equation}
\item  $Lks/N_f$ is the average pairwise Gaussian linking number \emph{per filament}, which approximates the average number of windings per filament and, in principle, could provide an approach to compare systems with different $N_f$,
    \begin{equation}
        \frac{Lks}{N_f} = \frac{1}{N_f}\sum_{j>i} \left| Lk_{i,j} \right| \approx \frac{N_w}{N_f};
    \end{equation}
\item  $\mathcal{E}$ is the average pairwise Gaussian linking number \emph{per filament pair}, which has been adopted in the main text and approximates the average number of windings per filament pair, thus acknowledging the pairwise nature of $Lk_{i,j}$,
    \begin{equation}\label{eq:GLNdef}
        \mathcal{E} = \frac{Lks}{N_p} = \frac{2}{N_f (N_f-1)}\sum_{j>i} \left| Lk_{i,j} \right| \approx \frac{N_w}{N_p}.
    \end{equation}
\end{itemize}

\begin{figure*}[h!]
\centering
\includegraphics[scale=0.715]{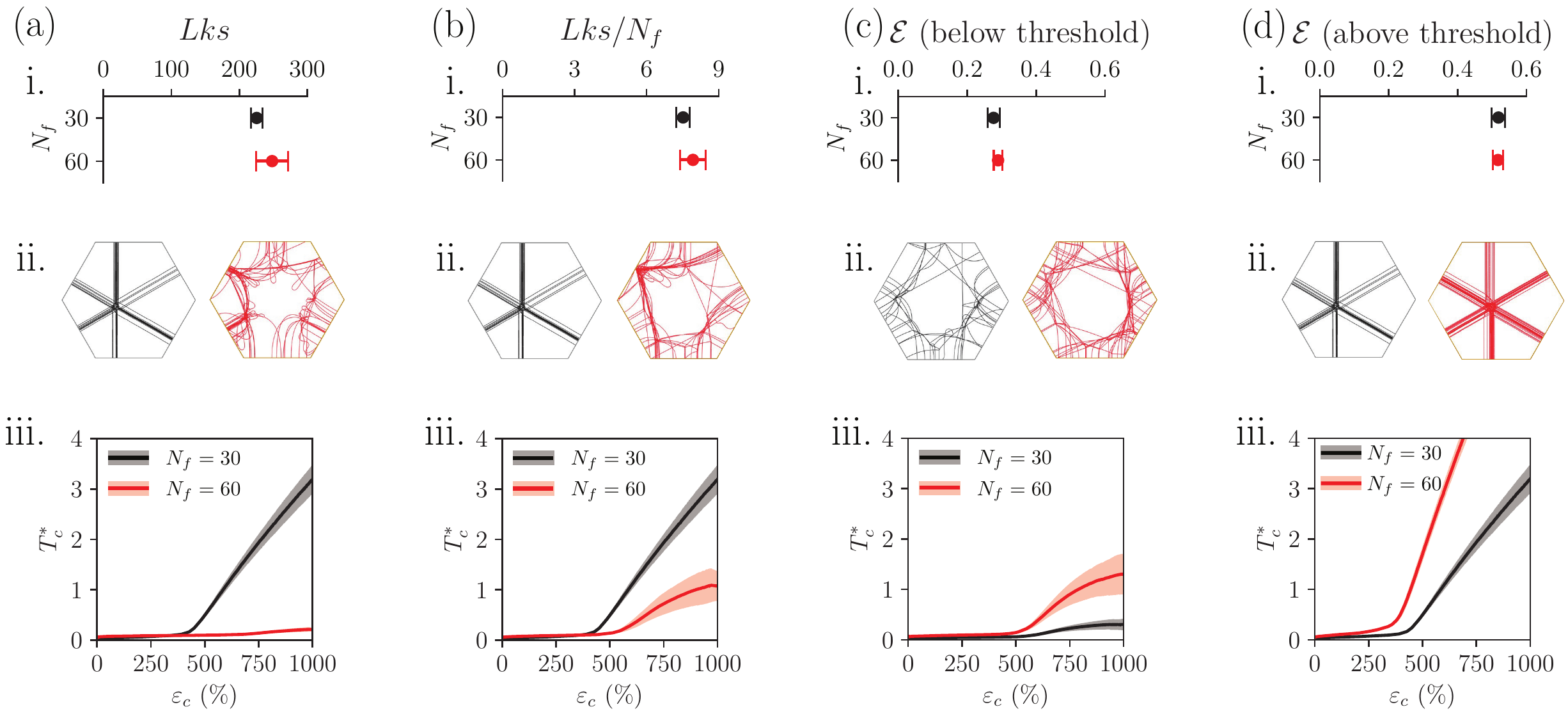}
\caption{\label{fig:FigS3} Link between alternative definitions of system-wide entanglement metric and self-organization, established by comparing filament configuration following stretch (ii) and mechanical response (iii) for networks with different number of filaments but comparable nominal entanglement (i) according to each considered metric (a-d). In (i), dot and braces denote mean $\pm$ standard deviation for $8$ model realizations. }
\end{figure*}

Neither $Lks$ nor $Lks/Nf$ are useful predictors of network self-organization, as evident when comparing ensembles with different $N_f$ and otherwise identical parameters  including the considered entanglement metric, Fig.~\ref{fig:FigS3}(a)-i, -ii and Fig.~\ref{fig:FigS3}(b)-i, -ii. Moreover, quantifying the emergent mechanical behavior of these networks also shows that, fixing $Lks$, Fig.~\ref{fig:FigS3}(a)-iii, or $Lks/N_f$, Fig.~\ref{fig:FigS3}(b)-iii, and increasing $N_f$ leads to a softer response. Thus, we discarded both of these measures of entanglement and adopted $\mathcal{E}$ in our study, since it correctly predicts self-organization regardless of $N_f$, Fig.~\ref{fig:FigS3}(c, d).

\clearpage

\section{\label{sec:entanglementVsStrain} Influence of cell deformation on the measure of  entanglement $\mathcal{E}$ }

To be meaningfully interpretable as a topological property of a network, $\mathcal{E}$ should not depend on reconfigurations of our filament ensemble that restrict filament crossings, such as those occurring during our stretching simulations with steric interactions. However, some dependence on deformation is expected because the pairwise linking is not a strict topological invariant for tangles. To test the robustness of the proposed entanglement metric, we report its variation during cell stretching simulations.

\begin{figure*}[h]
\centering
\includegraphics[scale=0.75]{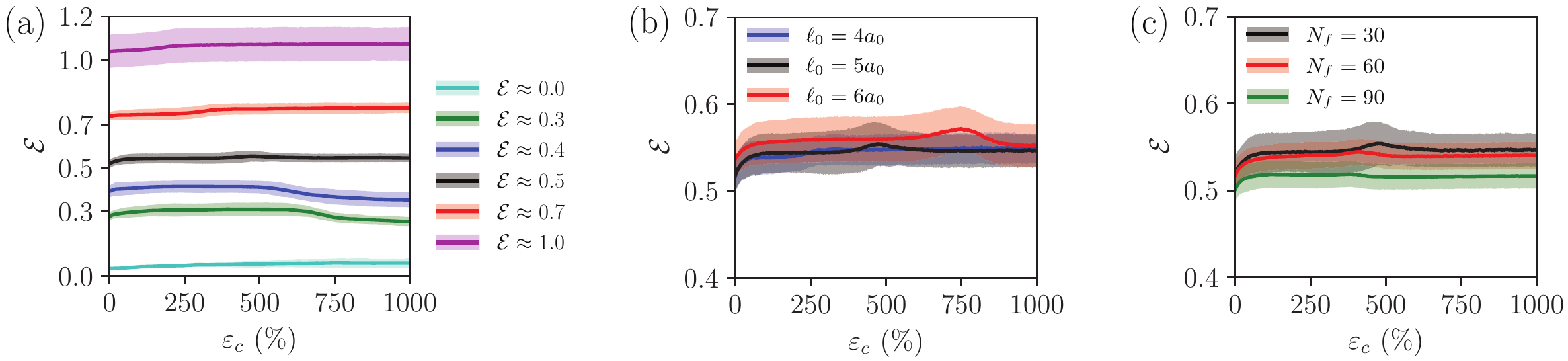}
\caption{\label{fig:FigS4} Evolution of network entanglement during stretching simulations for systems with clearly different $\mathcal{E}$ (a) and for systems with comparable entanglement but different $\ell_0$ (b) or $N_f$ (c). In all cases, $\mathcal{E}$ is very robust to deformations of the cell enclosing the filaments.  Solid lines and shadings: mean $\pm$ standard deviation for $8$ model realizations.}
\end{figure*}

For the systems compared in Fig.~\ref{fig:Fig2}(a), $\mathcal{E}$ is only mildly affected by the applied cell deformation and, despite not remaining constant, its variations during stretching do not modify the ranking of the networks according to their degree of entanglement, Fig.~\ref{fig:FigS4}(a).
Likewise, systems with different $\ell_0$, Fig.~\ref{fig:FigS4}(b), or $N_f$, Fig.~\ref{fig:FigS4}(c), but a comparable level of entanglement maintain similar and nearly constant values of $\mathcal{E}$ throughout stretching. Taken together, these data show that $\mathcal{E}$ is a robust measure of entanglement for our networks.

\clearpage

\section{\label{sec:AnalyticalModel} Derivation of an analytical model for equibiaxial cell stretching}

We aim to establish an analytical model  {explaining the emergent mechanical behavior of a network with entanglement above the topological threshold  in terms of its mechanical, geometrical, and topological properties. To this end}, we focus on a 2D projection of the  system on a plane that is parallel to the base of the prism enclosing $N_f$ {identical} filaments {with} constant elastic modulus, $E$, {reference length $\ell_0$, and} circular cross-sectional area $A_f = \pi \phi^2 / 4$. {Since all filaments are identical, application of an equibiaxial cell areal strain, $\varepsilon_c$, must result in all of them developing the same  strain, $\varepsilon_f = (\ell - \ell_0) / \ell_0$, leading to the following expression for} the total cell force:
\begin{equation}\label{eq:Eq1}
    F_c = \sum_{i=1}^{2 N_f}{F_i} = 2 N_f E A_f \varepsilon_f.
\end{equation}
Thus, the corresponding dimensionless nominal cell tension is:
\begin{equation}\label{eq:Eq1ten}
    T_c^* = \frac{F_c \, a_0}{EA_f N_e s_0} = \frac{N_f \, \varepsilon_f}{N_e \tan(\pi/N_e)},
\end{equation}
where we have used the geometrical relation between  the side and apothem length of a regular polygon, $s_0 = 2 a_0 \tan(\pi/N_e)$.

\begin{figure*}[b]
\centering
\includegraphics[scale=0.69]{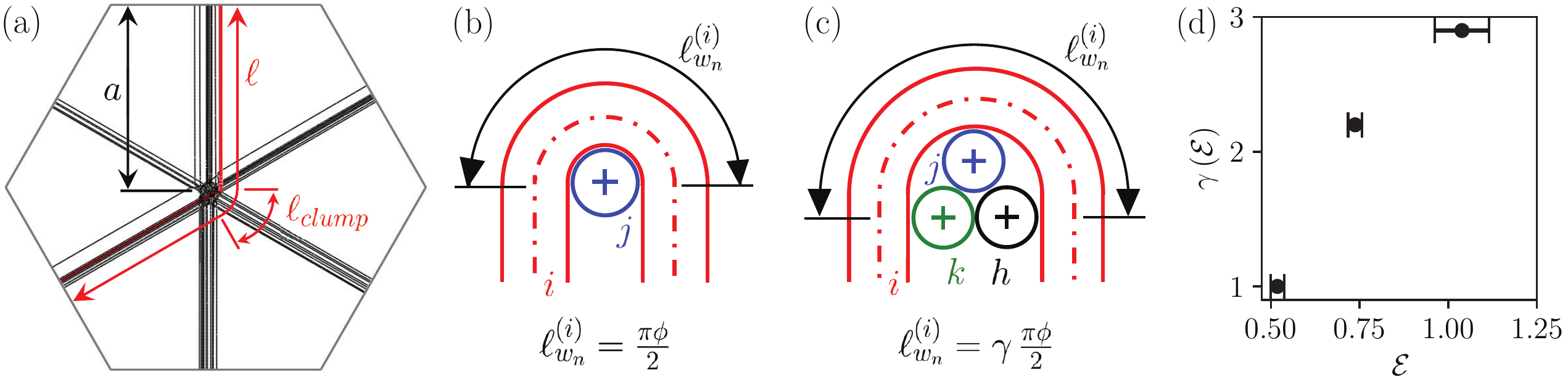}
\caption{\label{fig:FigS6} Derivation of a 1D analytical model {for equibiaxial} cell stretching. (a) {Sketch of the} filament {configuration beyond the activation strain for a network whose degree of entanglement exceeds the threshold for mechanical activation of all filaments, where t}he current filament length, $\ell$, and apothem length, $a$, are related via the length stored in the central tight tangle, $\ell_{clump}$. (b) Sketch of the simplest type of winding, formed by  filament $i$ looping around  filament $j$. (c) Windings that involve more filaments correspond to a larger filament length stored in the tight tangle, as captured by the scalar parameter $\gamma(\mathcal{E})\geq1$. {(d) Estimation of $\gamma$ as a function of $\mathcal{E}$, obtained by matching the analytical model to the average mechanical response of $8$ model realizations for each considered value of $\mathcal{E}$ (dot and braces: mean $\pm$ standard deviation).}}
\end{figure*}

To determine the filament strain, we need to link the cellular deformation to the current filament length. From simple kinematical arguments, we can express the current apothem length, $a$, in terms of its reference value, $a_0$, and the cell areal strain, $\varepsilon_c$:
\begin{equation}\label{eq:Eq2}
    a = a_0 \sqrt{\varepsilon_c + 1}.
\end{equation}
Assuming that the central tight tangle {emerging} at large cell deformations is located at the center of the enclosing regular polygon, we can approximately link the current apothem length {to} the current filament length, $\ell$, by {accounting for} the length stored in the tight tangle {formation}, $\ell_{clump}$, as sketched in Fig.~\ref{fig:FigS6}(a):
\begin{equation}\label{eq:Eq3}
    a \approx \frac{1}{2}\left(\ell - \ell_{clump}\right).
\end{equation}

To determine $\ell_{clump}$, we first consider the case of  filament $i$  being wound around filament $j$ in the simplest possible way, Fig.~\ref{fig:FigS6}(b). The length used to form such a winding is simply
\begin{equation}\label{eq:Eq4}
	\ell_{w_n}^{(i)} = \frac{\pi \phi}{2} = \bar{\ell}_w.
\end{equation}
{If} filament $i$ forms $N_w^{(i)}$ windings of this kind with other filaments in the ensemble, the length available to bridge the cell walls {is} reduced by the quantity
\begin{equation}\label{eq:Eq5}
    \ell_{clump} = N_w^{(i)} \bar{\ell}_w = \frac{N_w^{(i)} \pi \phi}{2}.
\end{equation}
However, filament {windings might be} more complex than the one sketched in Fig.~\ref{fig:FigS6}(b). For instance, a filament could wind around more than a single filament, Fig.~\ref{fig:FigS6}(c), such that the length $\ell_{w_n}^{(i)}$ {should} be larger than $\bar{\ell}_w$. To account for this extra length due to more complex windings, we modify Eq.~(\ref{eq:Eq5}) as 
\begin{equation}\label{eq:Eq8}
    \ell_{clump} = \gamma N_w^{(i)} \bar{\ell}_w = \frac{\gamma}{2} N_w^{(i)} \pi \phi.
\end{equation}
where $\gamma$ can be interpreted as a phenomenological measure of the typical complexity of the windings in a given system. {The} value {of $\gamma$ should} be $1$ when all windings correspond to {that} sketched in Fig.~\ref{fig:FigS6}(b), {and should} increase for more complex windings, Fig.~\ref{fig:FigS6}(c). 

Now, recalling the definition of $\mathcal{E}$, Eq.~(\ref{eq:GLNdef}), the number of windings formed by filament $i$ in a system {featuring} a total of $N_w$ windings {can be estimated} as
\begin{equation}\label{eq:Eq9}
    N_w^{(i)} \approx \frac{N_w}{N_f} \approx \frac{\mathcal{E}}{2} \left(N_f-1\right),
\end{equation}
and hence Eq.~(\ref{eq:Eq8}) {can be rewritten as}
\begin{equation}\label{eq:Eq10}
    \ell_{clump} \approx \frac{\gamma}{4} \mathcal{E} \left(N_f-1\right) \pi \phi.
\end{equation}
Replacing this expression into Eq.~(\ref{eq:Eq3}), we obtain an explicit relation between the current apothem and filament lengths, the network entanglement, and the number of filaments in the system:
\begin{equation}\label{eq:Eq11}
    a \approx \frac{1}{2} \left[\ell - \frac{\pi}{4} \gamma \, \mathcal{E} \phi \left(N_f-1\right) \right].
\end{equation}

Finally, we can relate the cell areal strain, $\varepsilon_c$, and the current filament length, $\ell$, by combining Eqs.~(\ref{eq:Eq2}, \ref{eq:Eq11}):
\begin{equation}\label{eq:Eq12}
    a_0 \sqrt{\varepsilon_c + 1} \approx \frac{1}{2} \left[\ell - \frac{\pi}{4} \gamma \, \mathcal{E} \phi \left(N_f-1\right) \right],
\end{equation}
so that
\begin{equation}\label{eq:Eq13}
    \ell \approx 2 a_0 \sqrt{\varepsilon_c + 1} + \frac{\pi}{4} \gamma \, \mathcal{E} \phi \left(N_f-1\right).
\end{equation}
The {above expression can be used to determine the areal strain at which filaments become taut and thus begin to  contribute to the cellular response, which we term \emph{cell activation strain}}, $\varepsilon_c^A$, {by setting $\ell = \ell_0$ in Eq.~(\ref{eq:Eq13}) and solving for $\varepsilon_c = \varepsilon_c^A$}:
\begin{equation}\label{eq:Eq14}
    \varepsilon_c^A \approx \frac{1}{4 a_0^2} \left[ \ell_0 - \frac{\pi}{4} \phi \, \gamma(\mathcal{E}) \, \mathcal{E} \left(N_f - 1\right) \right]^2 - 1.
\end{equation}

{For $\varepsilon_c < \varepsilon_c^A$, the filaments are slack, and hence their strain is $\varepsilon_f = 0$. On the other hand, beyond the activation strain, $\varepsilon_f = (\ell-\ell_0)/\ell_0$. Introducing the Macaulay brackets, defined by $\langle a \rangle$ = 0 if $a<0$ and $\langle a \rangle = a$ otherwise, we obtain:}
\begin{equation}\label{eq:Eq15}
    \varepsilon_f =  \left\langle \frac{\ell - \ell_0}{\ell_0} \right\rangle \approx \frac{2 a_0}{\ell_0} \left\langle \sqrt{\varepsilon_c + 1} - \sqrt{\varepsilon_c^A + 1} \; \right\rangle.
\end{equation}

{Lastly, substituting Eq.~(\ref{eq:Eq15}) into Eq.~(\ref{eq:Eq1ten}), we obtain the expression provided in Eq.~(\ref{eq:cellForce_an}) in the main text for the emergent dimensionless cell tension for a corralled and entangled filament ensemble subjected to equibiaxial stretch:}
\begin{equation}\label{eq:Eq16}
    T_c^* \approx \frac{2 N_f a_0}{\ell_0 N_e \tan(\pi/N_e)}     \left\langle\sqrt{\varepsilon_c + 1} - \sqrt{\varepsilon_c^A + 1} \; \right\rangle.
\end{equation}

\clearpage

{\section{\label{sec:DefRate} Frictional hindrance of  non-affine IF reorganization at high strain rate}}

As discussed in the main text, IF networks interact with other cellular structures, including actin filaments, microtubules, and the nucleus \cite{Huber2015,Kechagia2022.03.01.482474}. These interactions should hinder the massive reorganizations of IF networks identified here. Because IF turnover is amongst the slowest, we expect that these interaction will effectively result in a small frictional resistance if loading rate is slow enough. Here, we examine the situation in which these frictional interactions relative to an affinely deforming cytosolic medium  are not small. The rate-dependent forces that we derive next are different from the native frictional forces relative to a fixed background implemented in \emph{cyto\textbf{sim}}.

\begin{figure*}[h]
\centering
\includegraphics[scale=0.61]{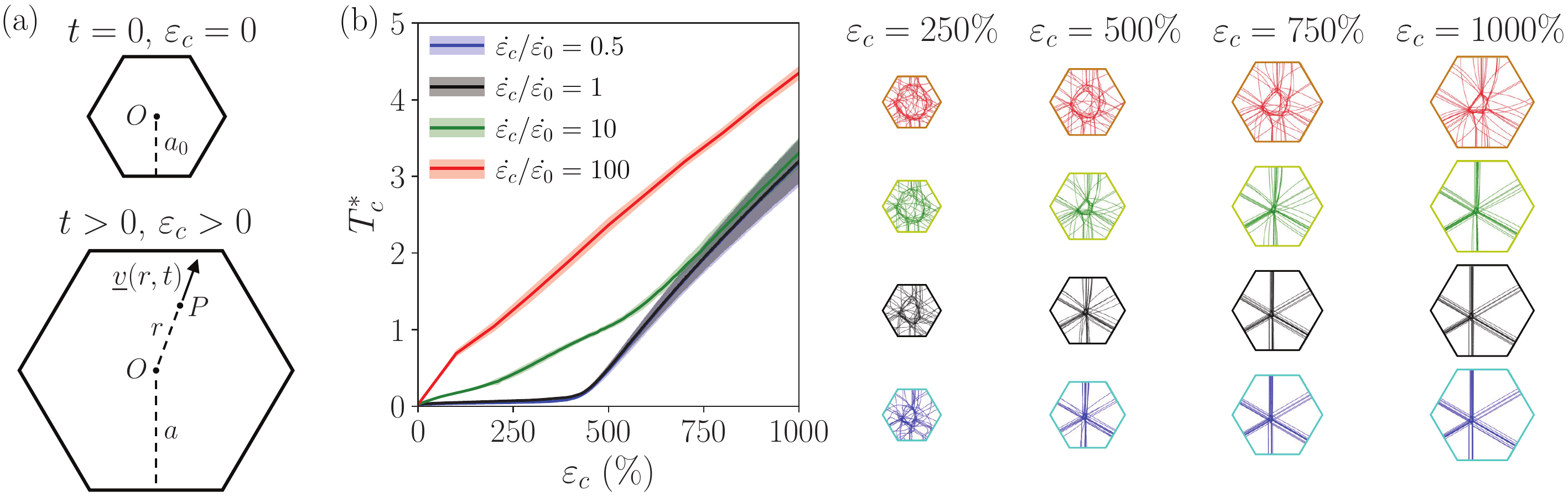}
{\caption{\label{fig:FigS9} Cell stretching accounting for the frictional interaction with an affinely deforming  cytosolic background. (a) Sketches supporting the derivation of the expression for the background velocity emerging from cytosolic expansion in response to rapid cell stretching. (b) Influence of applied strain rate on the mechanical response and self-organization of entangled networks. Solid lines and shadings: mean $\pm$ standard deviation of $8$ model realizations.}}
\end{figure*}

To determine the  drag forces induced by an affinely deforming background, we begin by considering a hexagon of reference apothem length $a_0$ and reference area $A_0$, Fig.~\ref{fig:FigS9}(a), which is deformed equibiaxially over a time period $t_{load}$ to reach a maximum apothem length $a_{max}$ and a maximum area $A_{max}$, corresponding to an areal strain $\varepsilon_{c,\,max} = \left(A_{max} - A_0 \right) / A_0$, with a constant strain rate $\dot{\varepsilon}_c = \varepsilon_{c,\,max}/t_{load}$. 
For simplicity, we assume that the hexagon center, $O$, remains fixed as the deformation is applied. Since the imposed stretch causes the cytosol to expand isotropically in the hexagon plane, a simple calculation shows that the velocity field of an affinely deforming cytosolic background  at a generic point located at a distance $r$ from the origin $O$ at time $t$ will be directed along the local radial direction with magnitude
\begin{equation}\label{eq:Eq20}
	v(r,t) = \frac{r}{t_{load}} \frac{ \sqrt{ 1 + \varepsilon_{c,\,max} } - 1 }{ \sqrt{1 +  \varepsilon_{c,\,max} \frac{t}{t_{load}}} }.
\end{equation} 
The magnitude of the corresponding drag force on the filaments associated with a mobility coefficient $\mu$ is $f(r,t) = {v(r,t)}/{\mu}$.

{Next, we report the results of cell stretching simulations performed at varying strain rates, where we account for the role of a background friction that idealizes the presence of other cytoskeletal components besides IF bundles. As shown in Fig.~\ref{fig:FigS9}(b), for sufficiently slow deformations the models exhibit the same characteristic non-affine and nonlinear response reported in the main text, \textit{cf.} black and blue curves, shadings, and models, confirming that the role of a frictional cytosolic environment can be neglected when the cell is stretched quasi-statically, in agreement with the results presented in the main figures for slow strain rates. Increasing the strain rate (green and red) leads to higher IF-induced cellular tension, a loss of the slack-taut nonlinear response, and an increasing affinity of filament deformation. These results are consistent with the IF-induced rate-dependent stiffening of cell monolayers reported in \cite{Duque2023.01.05.522736}, suggesting that this phenomenon is due to the hindrance posed by other cellular structures to non-affine IF reorganization.}

\clearpage

\section{\label{sec:Temperature} Influence of temperature on network stretching}

\begin{figure*}[h]
\centering
\includegraphics[scale=0.3]{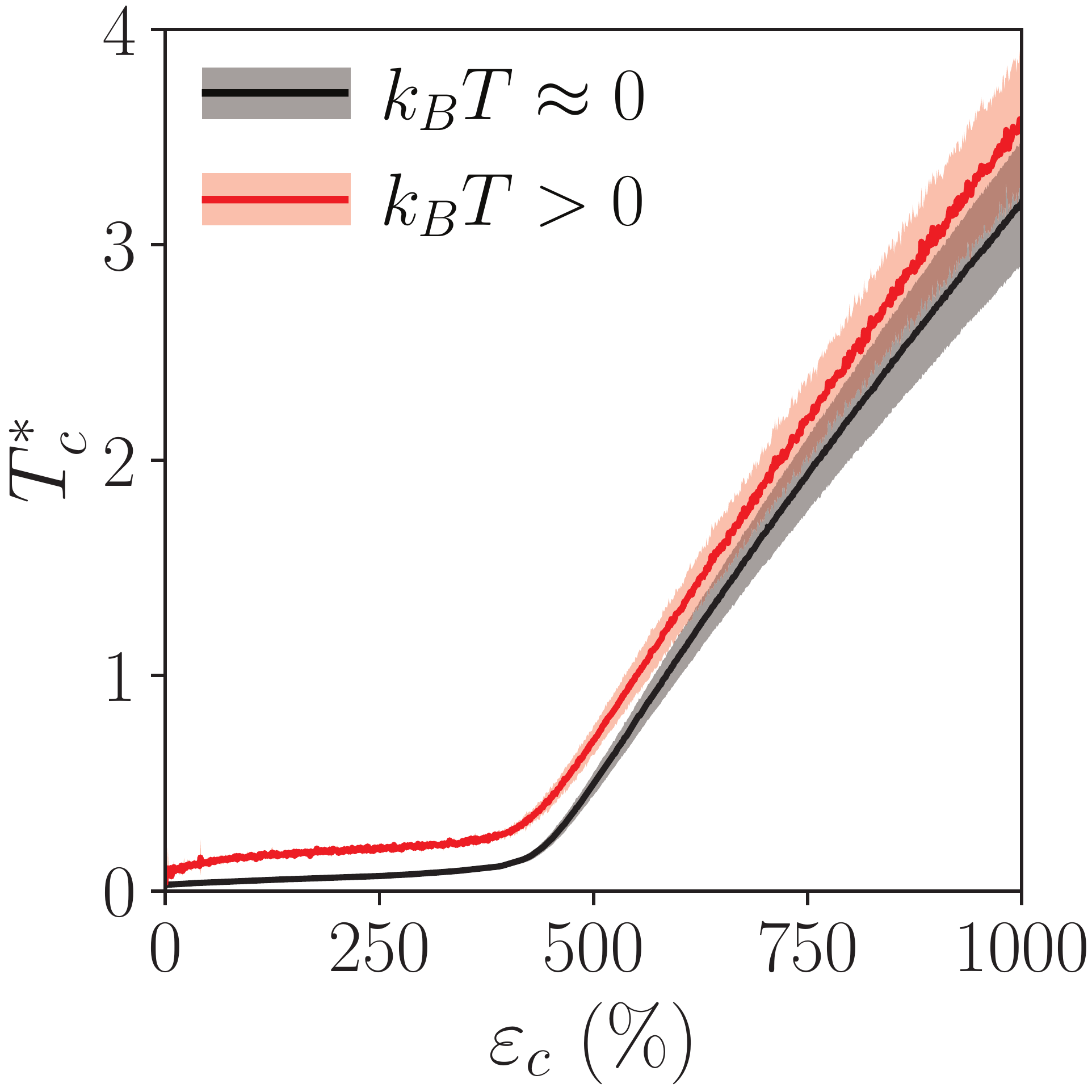}
\caption{\label{fig:FigS8} Influence of $k_B T$ on the mechanical response of entangled networks, confirming the negligible role of thermal fluctuations in the present context. Solid lines and shadings: mean $\pm$ standard deviation of $8$ model realizations.}
\end{figure*}

To validate the assumption that cell {stretching} can be {simulated athermally, as we did throughout our study, we compare thermal \textit{vs.} athermal} stretching for $8$ {model} realizations with default parameters.
{Since the results for these two cases differ only mildly,} Fig.~\ref{fig:FigS8}, neglecting the role of thermal fluctuations in the context of our study is a reasonable assumption.

\clearpage

\section{\label{sec:Bending} Influence of filament bending rigidity}

\begin{figure*}[h]
\centering
\includegraphics[scale=0.3]{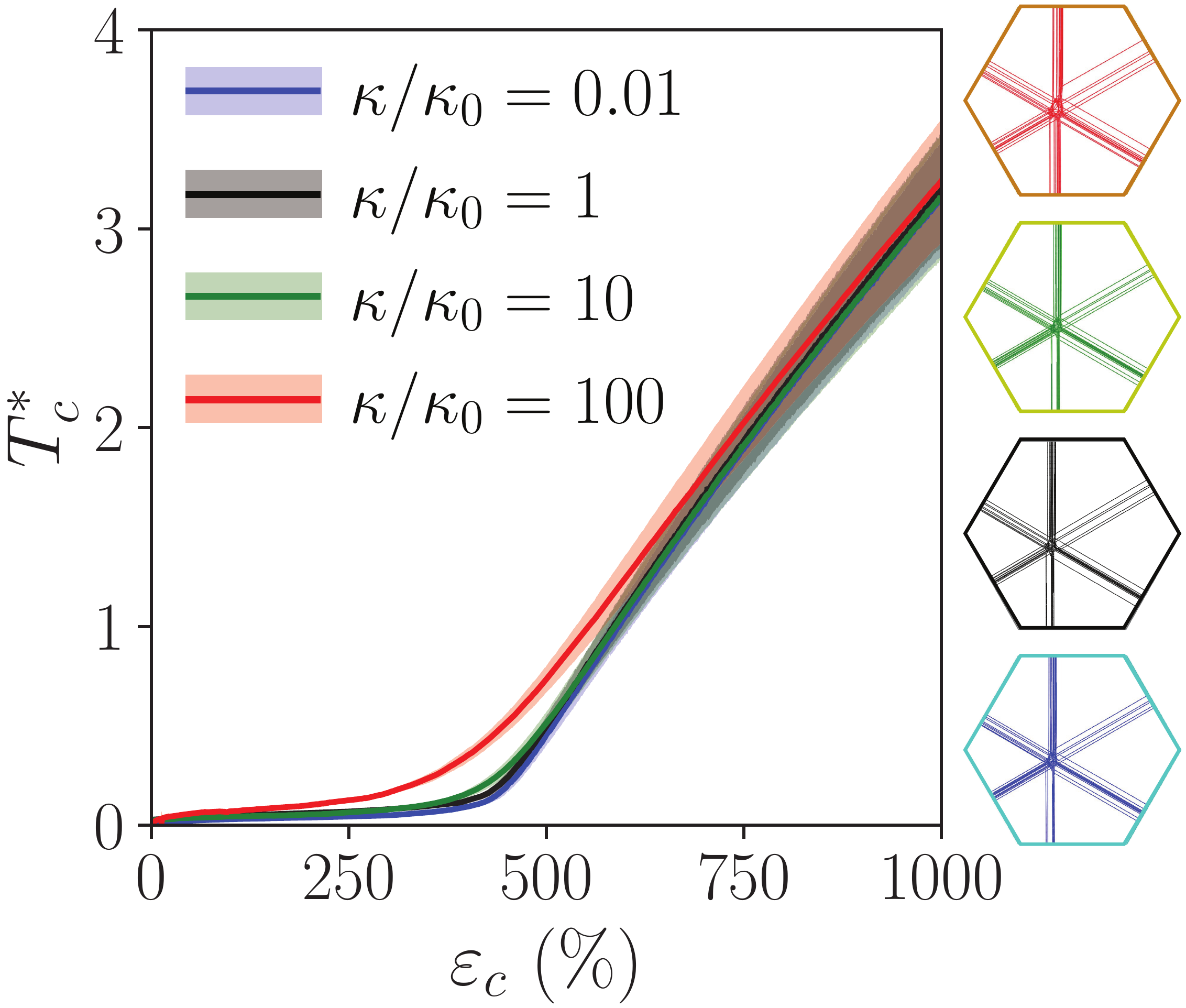}
\caption{\label{fig:FigS10} Influence of filament bending rigidity on the mechanical response {and self-organization} of entangled networks, confirming the negligible role of $\kappa$ within the examined range. Solid lines and shadings: mean $\pm$ standard deviation of $8$ model realizations.}
\end{figure*}

To confirm that the filament bending rigidity, $\kappa$, has a negligible role on the emergent mechanical response and network reorganization described in the main text, we simulate the stretching of $8$ {model} realizations where all parameters but $\kappa$ are set to their default values.
{Varying $\kappa$} in a range {spanning four} orders of magnitude {around the default value adopted in this study, $\kappa_0 = 0.021$ \si{\pico\newton\micro\meter\squared}}, we confirm its negligible influence on non-affine and nonlinear response of our networks, Fig.~\ref{fig:FigS10}, except for unreasonably large values of $\kappa$.

\clearpage

\section{\label{sec:Emodulus} Influence of filament elastic modulus}

\begin{figure*}[h]
\centering
\includegraphics[scale=0.55]{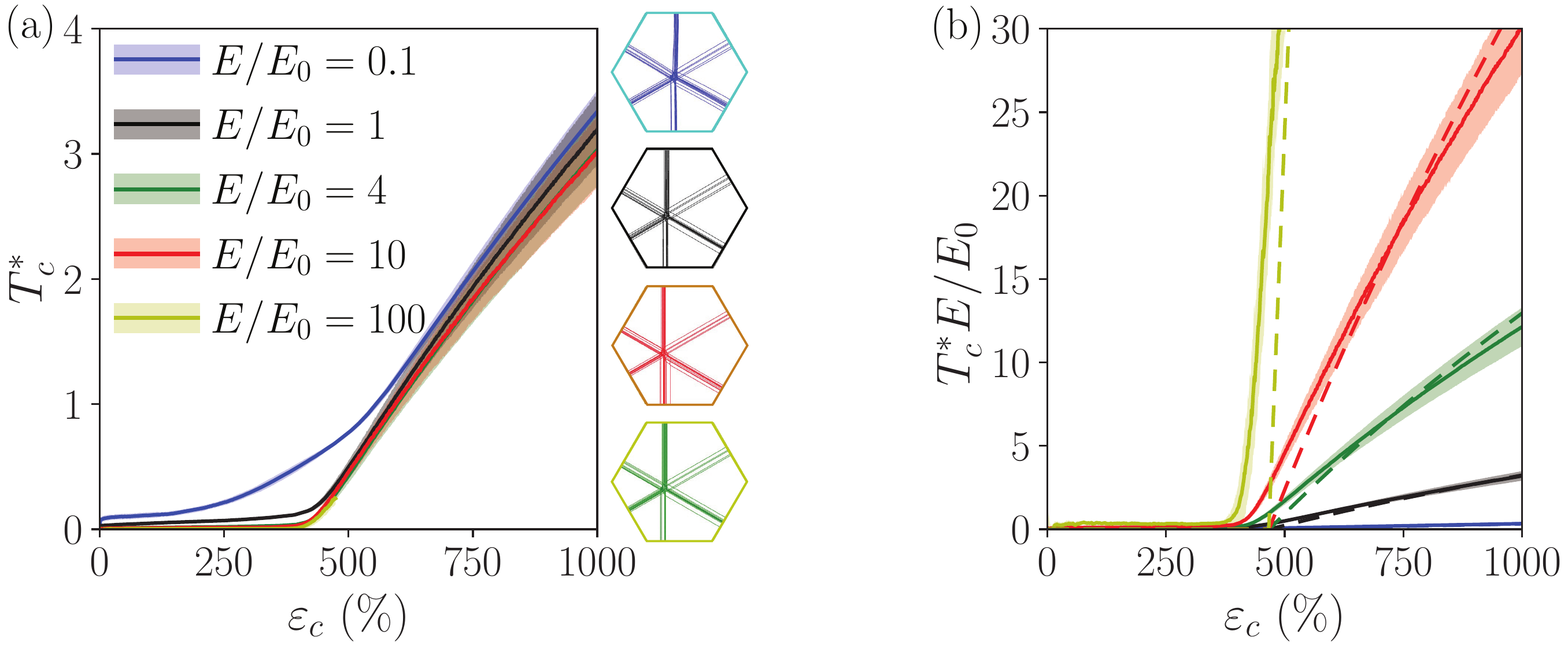}
\caption{\label{fig:FigS11} Influence of filament elastic modulus on the mechanical response {and self-organization} of entangled networks. Owing to the normalization included in its definition, $T_c^*$ is only marginally affected by $E$, {and mainly at moderate cell strains (a). This implies} that $T_c$ {should} depend linearly on $E$, as confirmed by representing the quantity $T_c^*E/E_0$ that corresponds to normalizing $T_c$ with respect to a fixed value of the filament elastic modulus (b). Solid lines and shadings: mean $\pm$ standard deviation of $8$ model realizations. Dashed line: 1D analytical model with $\gamma=1$.}
\end{figure*}

To address the influence of the filament elastic modulus $E$, we {simulate the stretching of $8$ {model} realizations where all parameters but $E$ are set to their default values. Varying $E$ in a range spanning three orders of magnitude around the default value adopted in this study, $E_0 = 13$ \si{\kilo\pascal}, we confirm its negligible influence on the network's ability to self-organize into star-shaped structures and on the \emph{dimensionless} cellular tension, particularly at large cell strains where the emergent response is dominated by filament stretching, Fig.~\ref{fig:FigS11}(a).
We note that the negligible influence of $E$ on $T_c^*$ follows from our definition of dimensionless tension, $T_c^* = T_c a_0 / EA_f$. Thus, the \emph{dimensional} cellular tension, $T_c = T_c^* EA_f / a_0$, should scale linearly with the filament elastic modulus. Representing the quantity $T_c^* E/E_0$, which corresponds to a normalization of $T_c$ with respect to a fixed value of the elastic modulus, confirms this expectation for both the simulations and the corresponding analytical model predictions, Fig.~\ref{fig:FigS11}(b).}

\newpage
\section{\label{sec:TransitionConstitutive} Influence of filament length distribution and constitutive behavior}

\begin{figure*}[h]
\centering
\includegraphics[scale=0.6]{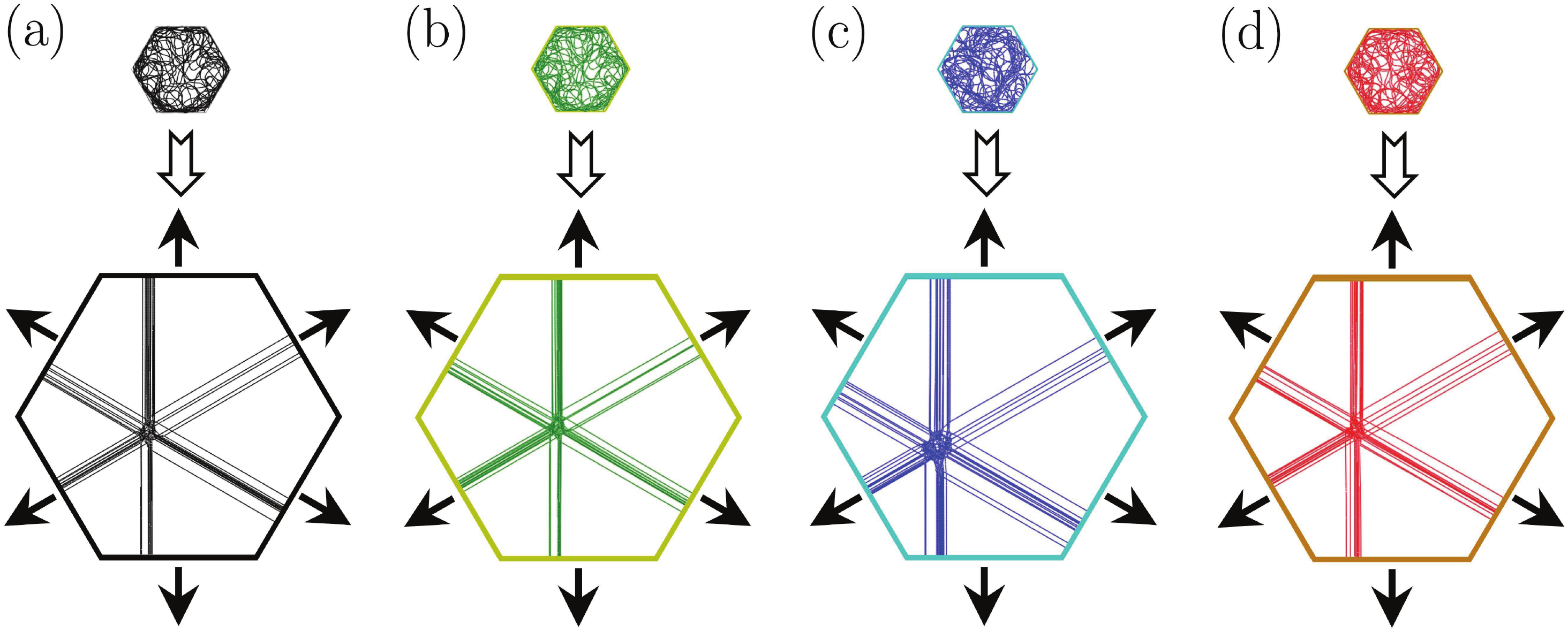}
\caption{\label{fig:FigS12} Slack-taut transition in {networks} that differ in terms of {filament} length distribution or constitutive behavior, showing that these parameters do not affect {non-affine network reorganization}. The models are colored according to the curves and shadings in Fig.~\ref{fig:Fig3}(c): (a) default model, where all filaments have equal initial length and are mechanically linear; (b) model featuring filaments with fixed initial length but a nonlinear constitutive behavior; (c) model comprising mechanically-linear filaments with a random initial length distribution; (d) model featuring nonlinear filaments with a random initial length distribution.}
\end{figure*}

As mentioned in the main text, {network} self-organization into star-shaped configurations is solely determined by {their degree of} entanglement, and should thus not depend on the specific mechanical response of individual filaments. Here, we {show the stretch-induced network reorganizations for the systems representative of those reported in Fig.~\ref{fig:Fig3}(c)}, which have varying  filament constitutive behavior, Fig.~\ref{fig:FigS12}(b, d), or  initial length distribution, Fig.~\ref{fig:FigS12}(c, d). Under equibiaxial loading, all models self-organize into similar characteristic star-shape configuration.

\newpage
\section{\label{sec:TransitionOtherShapes} Influence of enclosing cell shape}

\begin{figure*}[h]
\centering
\includegraphics[scale=0.3]{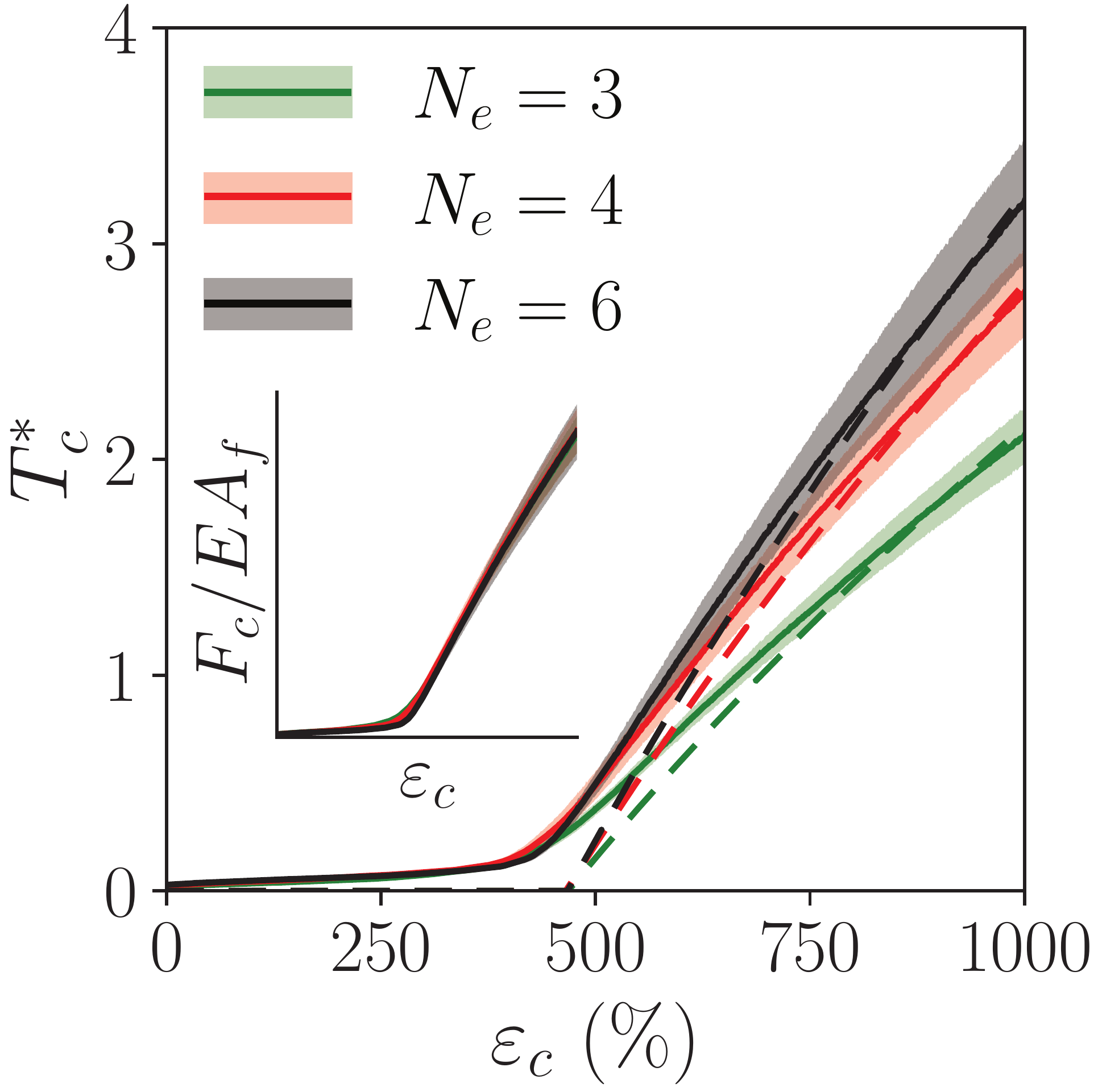}
\caption{\label{fig:FigS13} Influence of enclosing cell shape on the mechanical response of entangled networks. Since all networks have default parameters, the emergent response is only affected by the cell perimeter involved in the definition of $T_c^*$, which changes with $N_e$. The inset confirms that cellular forces are consistent across the considered models. Solid lines and shadings: mean $\pm$ standard deviation of $8$ model realizations. Dashed lines: 1D analytical model with $\gamma=1$.}
\end{figure*}

In the main text, we have shown that the occurrence of the slack-taut transition does not depend on the enclosing cell shape for triangular, square, or hexagonal prisms. Here, we focus on the emergent mechanical responses of $8$ model realizations with default parameters but enclosed in cells with $N_e = 3$, $4$, or $6$. While both our simulations and analytical predictions indicate larger $T_c^*$ for larger $N_e$, Fig.~\ref{fig:FigS13}, this effect is purely associated with the definition of $T_c^*$, which involves normalization of the cellular force by the reference cell perimeter, $2p_0 \propto N_e \tan(\pi / N_e)$, a quantity decreasing monotonically with $N_e$ for $N_e \geq 3$. Indeed, the cellular force is independent of $N_e$, as shown in the inset and also predicted analytically: $F_c/EA_f \approx 4 N_f a_0 \langle \sqrt{\varepsilon_c+1} - \sqrt{\varepsilon_c^A+1} \, \rangle / \ell_0$.

\begin{thebibliography}{60}
\bibitem [{sup()}]{supplement}%
  \BibitemOpen
  \href@noop {} {\bibinfo {title} {See {S}upplemental {M}aterial for further
  details on: model generation ({S}1), adopted model parameters ({S}2),
  quantification of pairwise linking number in discrete systems ({S}3), link
  between network entanglement metrics and emergent mechanics ({S}4), influence
  of cell deformation on $\mathcal{E}$ ({S}5), analytical model development
  ({S}6), role of other cytoskeletal structures towards hindering {IF}
  reorganization at high strain rates ({S}7), influence of system temperature
  ({S}8), filament bending rigidity ({S}9), elastic modulus ({S}10), length
  distribution and nonlinear constitutive behavior ({S}11), or enclosing cell
  shape ({S}12) on cell-scale mechanics.}}\BibitemShut {Stop}%
\end{thebibliography}
\end{document}